\documentclass[a4paper,fleqn,usenatbib]{mnras}

\usepackage{mathptmx}

\usepackage[T1]{fontenc}
\usepackage{ae,aecompl}

\usepackage{graphicx}	
\usepackage{amsmath}	
\usepackage{amssymb}	
\usepackage{wasysym}           
\usepackage{epstopdf}
\usepackage{mathrsfs}
\usepackage{natbib}
\usepackage{color}
\usepackage{hyperref}

\title{Tidal disruption by extreme mass ratio binaries and 
application to ASASSN-15lh}

\author[Coughlin \& Armitage]{
Eric R.~Coughlin$^{1}$\thanks{email: eric\_coughlin@berkeley.edu}\thanks{Einstein fellow}, 
Philip J.~Armitage$^{2,3}$
\\
$^{1}$Astronomy Department and Theoretical Astrophysics Center, University of California, Berkeley, Berkeley, CA 94720 \\
$^{2}$JILA, University of Colorado and National Institute of Standards and Technology, 440 UCB, Boulder, CO 80309-0440, USA \\
$^{3}$Department of Astrophysical and Planetary Sciences, University of Colorado, 391 UCB, Boulder, CO 80309-0391, USA \\
}

\date{Accepted XXX. Received YYY; in original form ZZZ}

\pubyear{2016}

\begin{document}
\label{firstpage}
\pagerange{\pageref{firstpage}--\pageref{lastpage}}
\maketitle

\begin{abstract}
Tidal disruption events (TDEs) observed in massive galaxies with inferred central black hole masses $M_h > 10^8 \ M_\odot$ are presumptive candidates for TDEs by lower mass secondaries in binary systems. We use hydrodynamic simulations to quantify the characteristics of such TDEs, focusing on extreme mass ratio binaries and mpc separations where the debris stream samples the binary potential. The simulations are initialised with disruption trajectories from 3-body integrations of stars with parabolic orbits with respect to the binary center of mass. The most common outcome is found to be the formation of an unbound debris stream, with either weak late-time accretion or no accretion at all. A substantial fraction of streams remain bound, however, and these commonly yield structured fallback rate curves that exhibit multiple peaks or sharp drops. We apply our results to the superluminous supernova candidate ASASSN-15lh and show that its features, including its anomalous rebrightening at $\sim 100$ days after detection, are consistent with the tidal disruption of a star by a supermassive black hole in a binary system. 
\end{abstract}




\section{Introduction}
Galaxy mergers lead to the formation of supermassive black hole (SMBH) binaries, which coalesce due to stellar scattering, interactions with nuclear gas, and gravitational wave losses (e.g., \citealt{begelman80,kelley17}). Although this scenario is theoretically compelling, observational tests have been frustrated by the difficulty of identifying  galaxies with multiple SMBHs. Imaging and spectroscopic surveys provide powerful tools for identifying samples of relatively wide separation dual Active Galactic Nuclei \citep{komossa03,comerford09}, but the number of confirmed binaries at pc-scale separation is much smaller \citep{rodriguez06}.  At even smaller separations, surveys of AGN variability have the potential to discover binaries approaching or entering the gravitational wave dominated phase of inspiral \citep{hayasaki07,macfadyen08,cuadra09,dorazio17}, with recent surveys identifying a number of candidates \citep{liu16,charisi16}.

Tidal disruption events \citep[TDEs;][]{lacy82,rees88}---when a star is destroyed by the tidal field of a supermassive black hole (SMBH)---offer a complementary route to SMBH binary discovery. While the number of observed TDEs is presently on the order of dozens \citep{komossa15}, current surveys such as the Panoramic Survey Telescope and Rapid Response System (Pan-STARRS; \citealt{kaiser10,chambers16,gezari12}), the All-Sky Automated Survey for SuperNovae (ASSASN; \citealt{shappee14,holoien14}), and the Palomar Transient Factory \citep{law09,arcavi14,blagorodnova17} are discovering events at an increasing rate. Future surveys such as the Zwicky Transient Facility \citep{bellm14} and the Large Synoptic Survey Telescope \citep{ivezic08} are predicted to yield hundreds of new candidates. Some fraction of these TDEs will occur in galaxies containing binaries, which may affect the light curve in distinct and predictable ways.

The detectability of SMBH binaries from TDE surveys is a strong function of the binary mass and separation. A binary modifies the fallback rate from the standard result for a single black hole, $\dot{M}_{\rm fb} \propto t^{-5/3}$ \citep{phinney89}, if the bound debris has an orbit that is large enough to experience a significant perturbation from the binary. For moderate mass ratio binaries with $q = M_2/M_1 \sim 0.2$, $M_2 \le M_1$ being the masses of the SMBHs, this requirement favours systems with relatively low mass SMBHs ($M_h \lesssim 10^6 \ M_\odot$) \citep{coughlin17}. More massive SMBH binary systems would still perturb the fallback rate, but only for separations where the gravitational wave inspiral time is short and the probability of a coincident TDE small.

\citet{coughlin17} used a combination of 3-body integrations and hydrodynamic simulations to identify the observable characteristics of TDEs in moderate mass ratio binary systems. Their results suggested that a binary of the appropriate separation can introduce a variety of strong perturbations to the light curve, including ``dips'' when fallback onto the disrupting hole is interrupted \citep{liu09,ricarte16}, bursts of accretion, and quasi-periodic variability. In this paper, we extend our prior study to the case of extreme mass ratio SMBH binaries. Our motivations are two-fold. From a theoretical perspective, disruptions by the secondary represent the {\em only} channel for TDEs in large galaxies where the primary has a mass $M_1 \gtrsim 10^8 \ M_\odot$ that is too large to tidally disrupt Solar-type (and smaller) stars. Minor mergers could result in a significant population of extreme mass ratio binaries in such galaxies, and because the gravitational wave inspiral time increases for a fixed primary mass approximately as $\tau_{GW} \propto 1/q$ \citep{peters63}, TDEs by the secondary could be perturbed by the presence of a primary  on observable time scales. Observationally, we investigate whether a TDE by the secondary in an extreme mass ratio system provides a possible interpretation for the light curve of the event ASASSN-15lh. Initially classified as an extremely luminous supernova \citep{dong16}, the physical nature of ASASSN-15lh remains unclear, with different lines of evidence supporting either a supernova \citep{godoy17} or TDE \citep{margutti17} interpretation. If ASASSN-15lh was a TDE, the likely mass of the primary SMBH in the observed galaxy is close to the maximum allowable value, leading to suggestions that the black hole is most likely rotating rapidly \citep{leloudas16}. Here we study whether the characteristics of the event---including its perplexing rebrightening at $\sim 100$ days after detection---could instead be consistent with the tidal disruption of a star by a binary companion with much lower mass.

The layout of the paper is as follows. In Section \ref{sec:timescales} we provide the basic timescales over which we expect variability to occur in TDEs resulting from extreme-$q$ SMBH binaries. We use our arguments to constrain the properties of the putative binary system that could explain ASASSN-15lh. To substantiate these general arguments, in Section \ref{sec:simulations} we describe the results of a number of hydrodynamic simulations of the disruption of a star by a SMBH binary, with binary properties appropriate to those anticipated for ASASSN-15lh. In Section \ref{sec:discussion} we discuss the implications of our findings, and we summarize and conclude in Section \ref{sec:summary}. 

\section{General Timescales}
\label{sec:timescales}
When the star is disrupted by a SMBH, there is a finite amount of time that passes before the debris stream returns to pericenter and starts accreting. Assuming that the tidal force acts impulsively and, hence, that the gas parcels follow ballistic orbits after the stellar center of mass passes through the tidal radius, this timescale is \citep{lodato09,coughlin14}

\begin{equation}
T_{ret} = \left(\frac{R_*}{2}\right)^{3/2}\frac{2\pi M_h}{M_*\sqrt{GM_h}}, \label{tr}
\end{equation}
where $R_*$ and $M_*$ are the stellar radius and mass and $M_h$ is the SMBH mass. Thereafter the accretion rate rises and reaches a peak in a time $T_{rise} = N\times{}T_{ret}$, where $N$ is a numerical factor of order unity that depends on the stellar composition \citep{lodato09} (there is very little dependence on the pericenter distance of the star, provided that it is within the tidal radius; \citealt{stone13,guillochon13}). 

For an isolated SMBH, the accretion rate eventually transitions to a power-law that is well-matched by the analytically-predicted, $t^{-5/3}$ scaling \citep{phinney89}\footnote{This may differ if the disruption is only partial \citep{guillochon13, mainetti17}, if the star is tightly-bound to the SMBH \citep{hayasaki13}, or if self-gravity results in the fragmentation of the stream \citep{coughlin15,coughlin16}, though we will not consider these possible complexities here}. Matters can be considerably more complicated when the disrupting black hole is part of a binary system \citep{coughlin17}; the accretion rate onto one or both SMBHs can often increase or decrease by orders of magnitude after the start of the decline. These changes to the accretion rate can also be quite erratic, and occur on timescales that are much shorter than the orbital period of the binary (though long-term accretion curves generally exhibit peaks in their power spectra at frequencies comparable to the orbital period). Moreover, in some extreme cases the accretion rate never follows the $t^{-5/3}$ fallback rate---even after a large number of $T_{ret}$. 
 
In spite of these complexities,  we can estimate the timescale over which one expects to \emph{start} to see variations in the accretion rate that differ from the analytic prediction for a single SMBH. In almost all cases the binary separation is much larger than the tidal radius for either black hole, and hence accretion onto the disrupting hole resembles that for an isolated SMBH at early times. (This only fails to be true for {\em extremely} tight binaries whose gravitational wave inspiral time is short.) At later times, however, the apocenter distance of the returning debris increases. As the apocenter distance approaches and exceeds the Roche lobe of the disrupting hole, the debris experiences a strong perturbation from the binary companion, its orbital properties change, and there will be a variation in the accretion rate evolution. Quantitatively, we assume a circular binary of semi-major axis $a$, black hole masses $M_1$ and $M_2 \leq M_1$, and define the mass ratio $q=M_2 / M_1$. For small $q$ the Roche lobe can be approximated as the Hill sphere radius,
\begin{equation}
 r_h = \left( \frac{q}{3} \right)^{1/3} a.
\end{equation} 
Assuming that the debris stream from a secondary disruption follows orbits with eccentricity $e \simeq 1$, and that it is significantly perturbed once it reaches an apocenter distance $d \gtrsim \epsilon r_h$, we expect deviations in the accretion rate after a time,
\begin{eqnarray}
 T_\sigma & \approx & \frac{\pi \epsilon^{3/2}}{\sqrt{6}} \frac{a^{3/2}}{\sqrt{GM_1}} \nonumber \\
 & \approx & 22 \epsilon^{3/2} \left( \frac{M_1}{10^8 \ M_\odot} \right)^{-1/2} \left( \frac{a}{\rm mpc} \right)^{3/2} \ {\rm days}.
\label{tsig} 
\end{eqnarray} 
Up to numerical factors this agrees with the result given for general $q$ by \citet{coughlin17}. There is no dependence on the mass of the disrupting, secondary SMBH.

If we hypothesize that ASASSN-15lh arises from the tidal disruption of a star by a SMBH binary, the above arguments allow us to estimate plausible parameters for the system. We first observe that the mass-luminosity relation applied to the host galaxy yields $M_h \simeq 10^8 \ M_{\odot}$ \citep{mcconnell13,leloudas16}, close to or beyond (depending upon the spin) the maximum possible mass for a TDE host. If the system is a binary, a disruption by the secondary is a simpler hypothesis. The basic timescales that constrain the parameters of the system are $T_{rise} \simeq 40$ days and $T_\sigma - T_{ret} \simeq $120 days, both of which are measured relative to the first observation (which we assume coincides with $T_{ret}$, though this depends on the efficiency of circularization; \citealt{guillochon15}). Letting the disrupted star be Solar-like and have a polytropic density profile with $\gamma = 5/3$ results in a peak time of $T_{peak} \simeq 1.6 T_r $ measured from the time that accretion begins. With $R_* = R_{\odot}$, $M_* = M_{\odot}$, $T_{peak} = 40$ d, and $M_h = M_2$ in Equation \eqref{tr}, we find 
\begin{equation}
M_2 \simeq 5\times10^{5}M_{\odot}. 
\end{equation}
Similarly, setting $T_{\sigma} = 120+T_{ret} \simeq 150$ days and $M_1 = 10^8M_{\odot}$ in Equation \eqref{tsig} gives (for $\epsilon=1$)
\begin{equation}
a \simeq 3.6 \text{ mpc}.
\end{equation}
Under the binary interpretation for the origin of ASASSN-15lh, the general scaling arguments suggest that the disrupting, secondary SMBH has a relatively low mass and is on a fairly tight orbit, with a period on the order of a year. For our hydrodynamic simulations we use the above estimates as a baseline by considering systems with $M_1 = 10^8 \ M_\odot$, $M_2 = 5 \times 10^5 \ M_\odot$, and separations of either 2.5~mpc or 5~mpc.

\section{Hydrodynamic Simulations}
\label{sec:simulations}
To test the reasoning outlined in the previous section, we simulate, using the smoothed-particle hydrodynamics (SPH) code {\sc phantom} \citep{price17}, the tidal disruption of a star by a SMBH binary. 

\subsection{Setup}
The star is approximated as a $\gamma = 5/3$ polytrope with a Solar mass and radius by first placing $\sim 5\times10^5$ particles on a close-packed sphere. The sphere is then stretched to achieve a nearly-polytropic density distribution, and it is thereafter dynamically relaxed for ten sound crossing times to smooth out any numerical perturbations. The relaxed polytrope is subsequently placed on an orbit that will take it within the tidal radius of a $5\times10^5M_{\odot}$ SMBH, which is the secondary SMBH in a circular binary with a primary of mass $M_1 = 10^8 \ M_\odot$. We consider cases where the semimajor axis of the binary orbit is $a = 500GM_1/c^2 \simeq 2.5$~mpc (30 simulations), and $a = 5$~mpc (20 simulations). 

To further establish the orbit of the to-be-disrupted star we use the same procedure outlined in \citet{coughlin17}: we assume that the star originally comes from near or beyond the sphere of influence of the binary on a parabolic orbit about the binary center of mass, its initial position randomly distributed over a sphere of radius $50a$ and its pericenter uniformly distributed between 0 and $2a$. We integrate the orbit of the star, assumed to be a point mass, about the binary until it is either ejected (recedes to beyond 100 times the separation of the binary) or passes through the tidal radius of either SMBH. If a star is disrupted by the secondary, we trace its orbit back to the point where it was five times the tidal radius away from the hole -- just prior to disruption -- and use its location and velocity at that time to initialize the SPH simulation.

By following this route we avoid simulating the hydrodynamics of the entire encounter of the star with the binary, which would be prohibitively expensive and would not yield much new information about the evolution of the star prior to disruption. As pointed out by \citet{coughlin17}, this method also highlights the fact that the orbital energy and angular momentum of the star can change as it orbits the binary. These changes can then impact the evolution of the TDE, and can introduce some intrinsic scatter to, e.g., the time to peak. As was also done in \citet{coughlin17}, we do not simulate any encounters that have $\beta = r_t/r_p \ge 5$, where $r_p$ is the pericenter distance of the star and $r_t$ is the tidal radius, as these encounters are difficult to resolve and can introduce numerical artifacts that affect the energy distribution of the debris. 

The evolution of the tidally-disrupted debris is followed for roughly an orbital period of the binary, which is on the order of a year, and self-gravity is included at all stages using a k-D tree \citep{gafton11}; the gas maintains a polytropic equation of state, and so cooling and heating through shocks are ignored. Particles that enter within 0.5$r_t$ of either black hole, with $r_t$ appropriately scaled to the size of the hole, are ``accreted,'' and contribute to the accretion rate of the black hole. We assessed the sensitivity of our accretion rates to this choice of accretion radius by decreasing it by a factor of two, and this resulted in only small differences (see Figure \ref{fig:mdots_comps}).  

\begin{figure*} 
   \centering
   \includegraphics[width=0.495\textwidth]{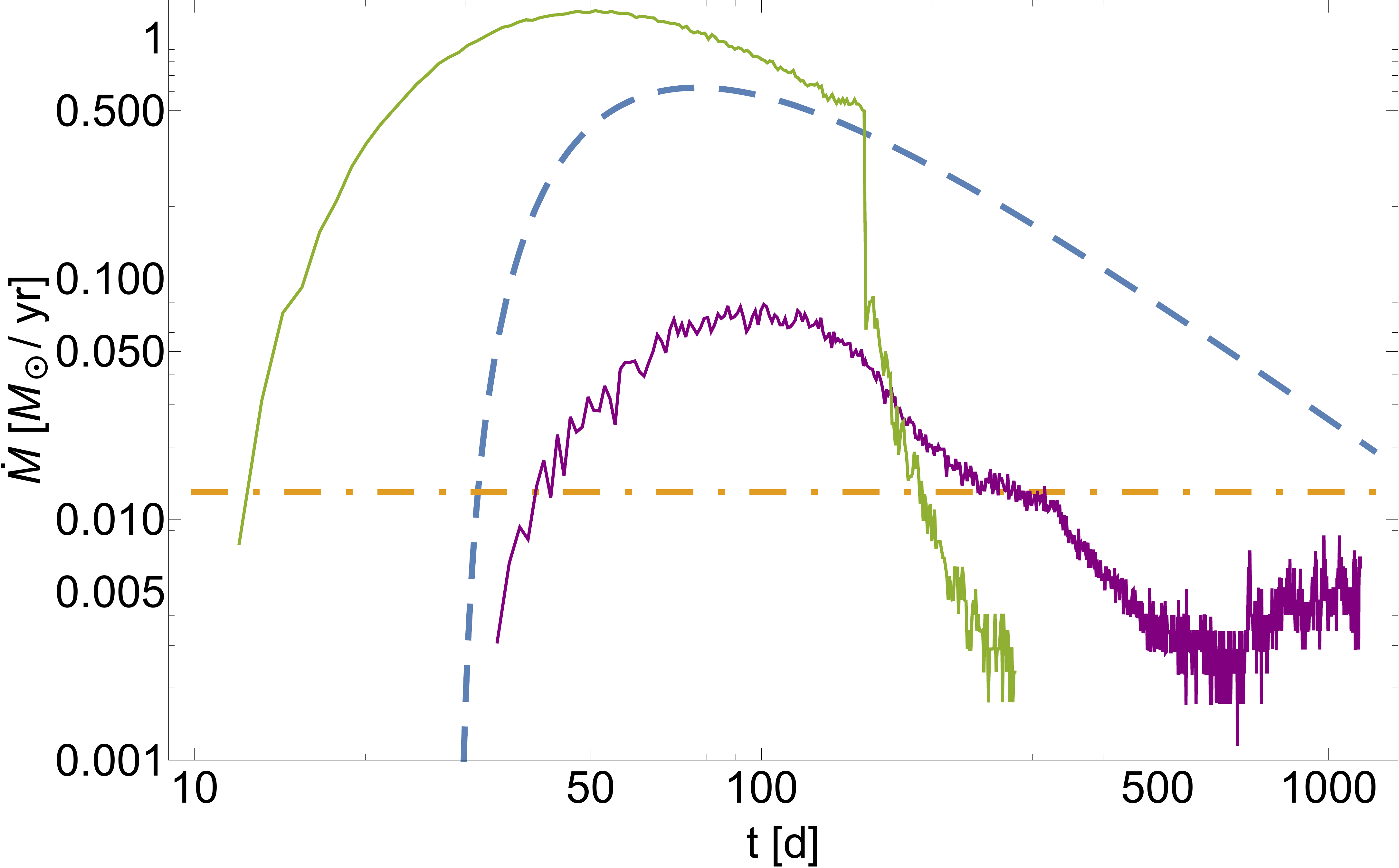} 
   \includegraphics[width=0.495\textwidth]{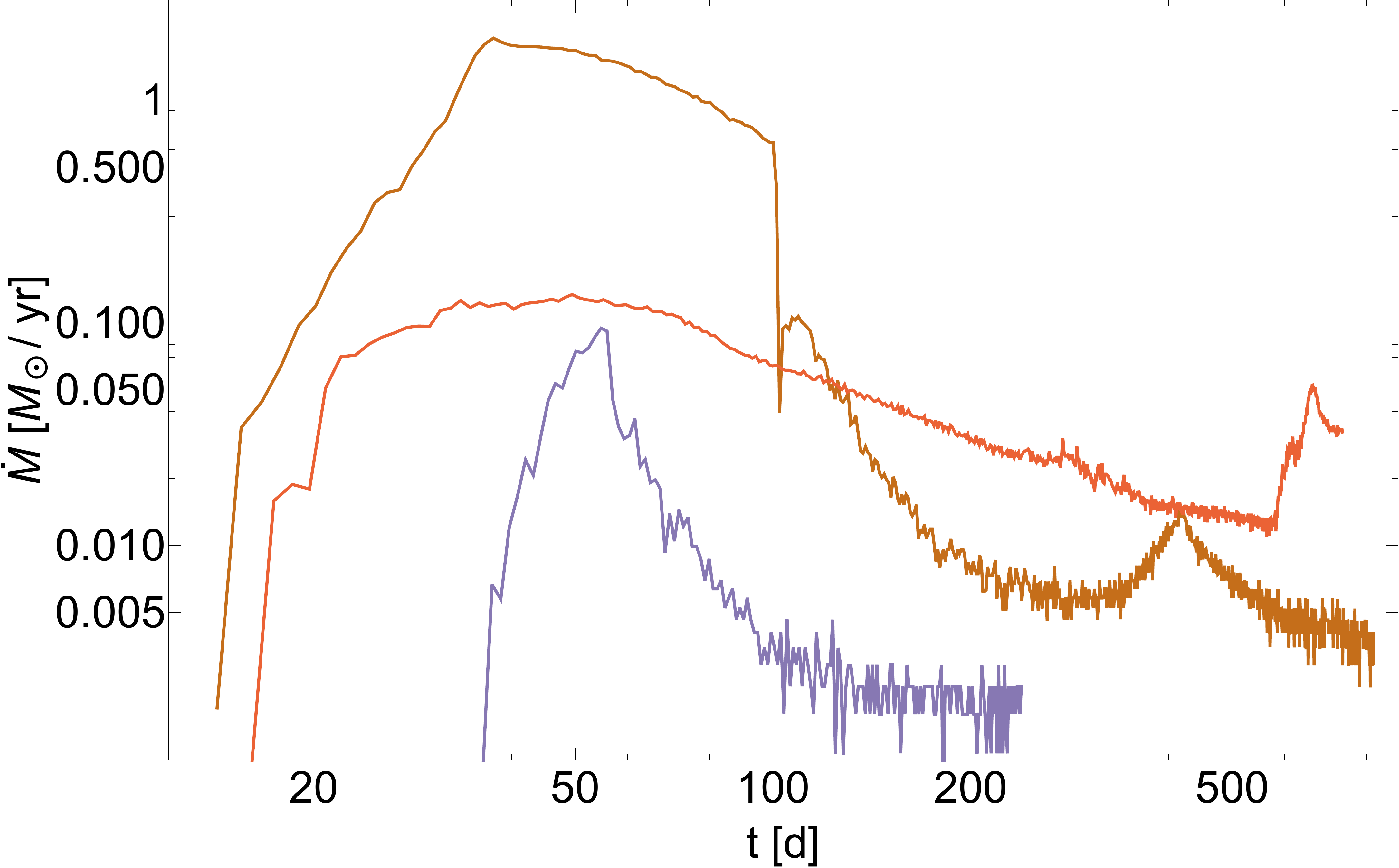} 
   \includegraphics[width=0.495\textwidth]{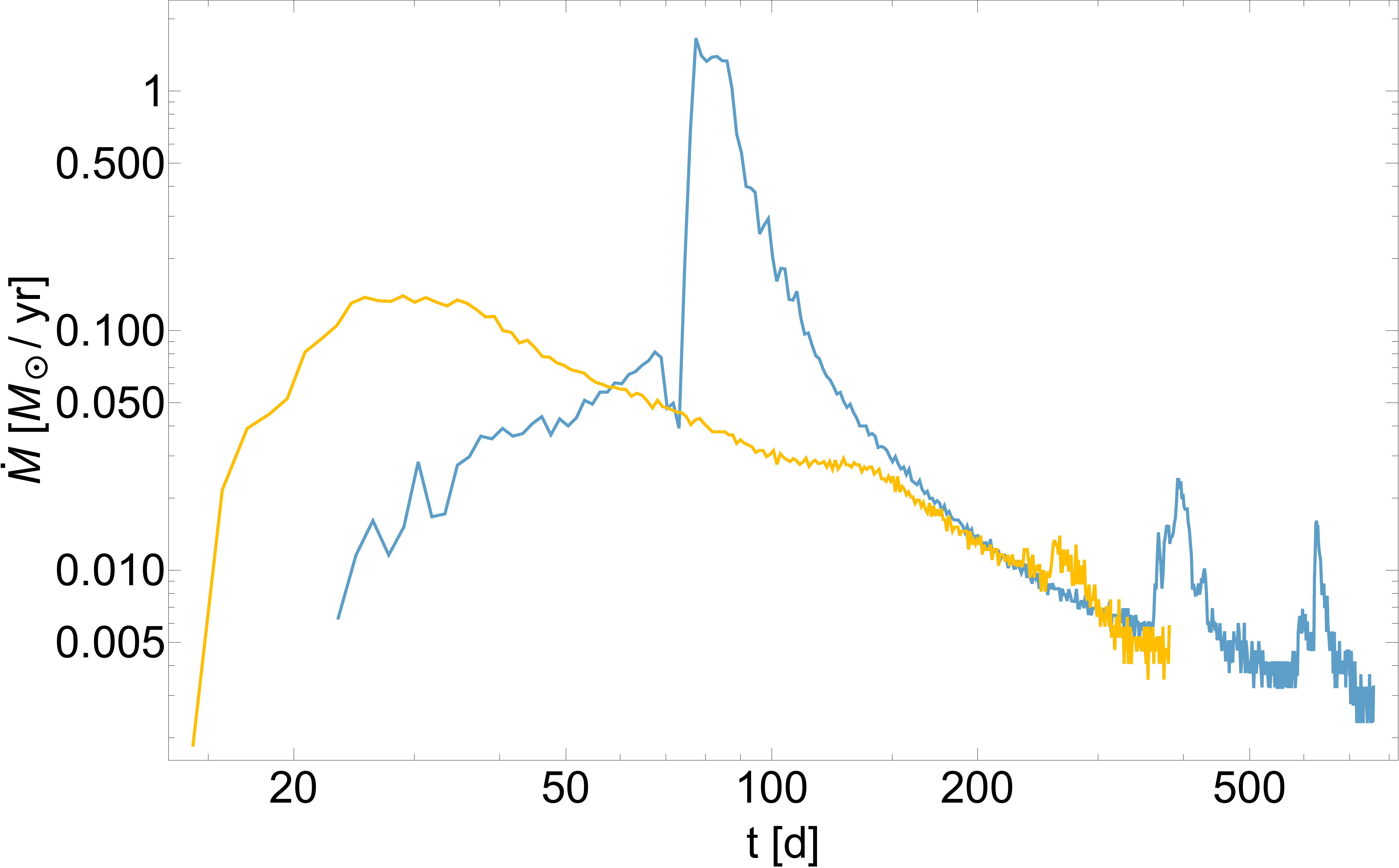} 
      \includegraphics[width=0.495\textwidth]{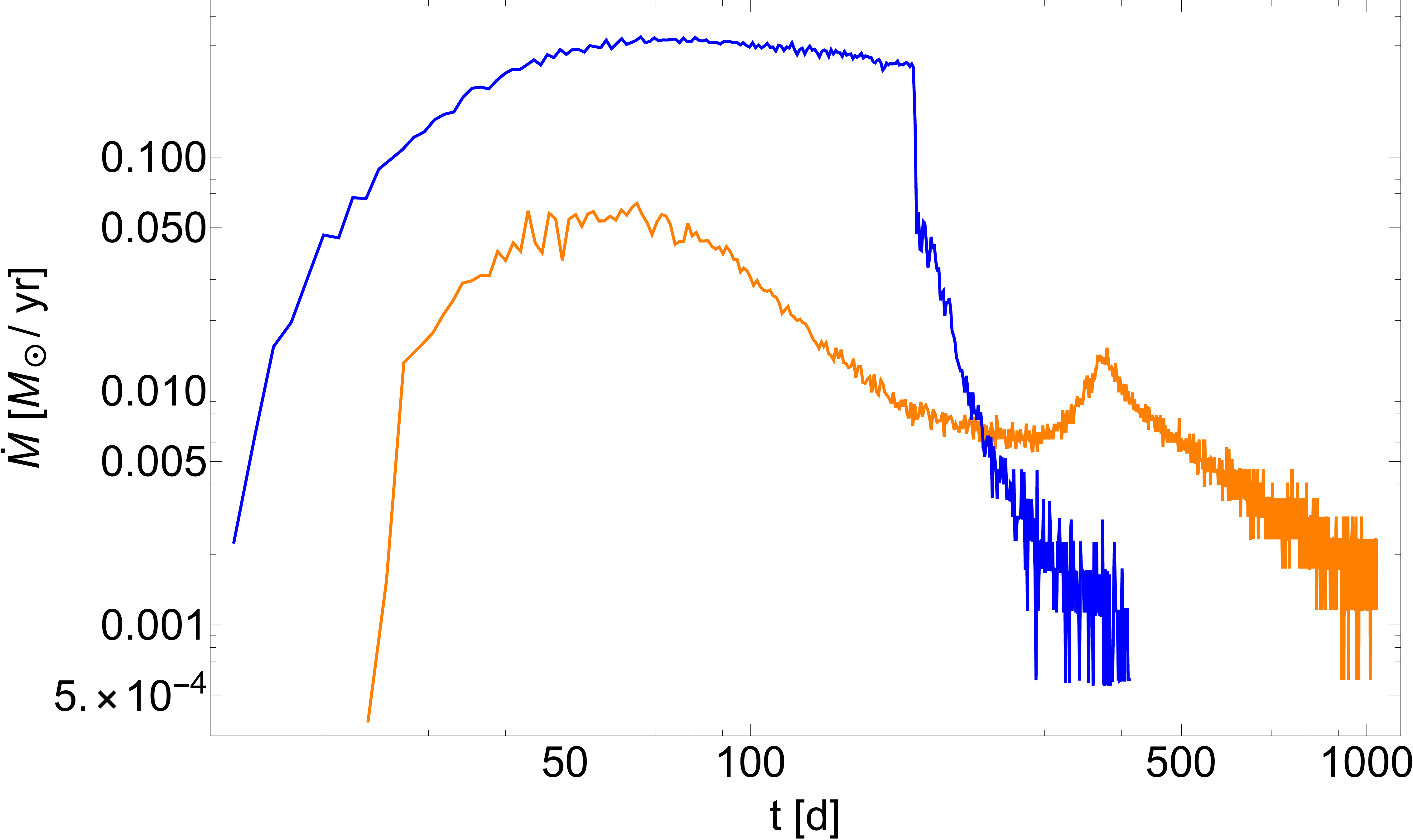} 
   \includegraphics[width=0.495\textwidth]{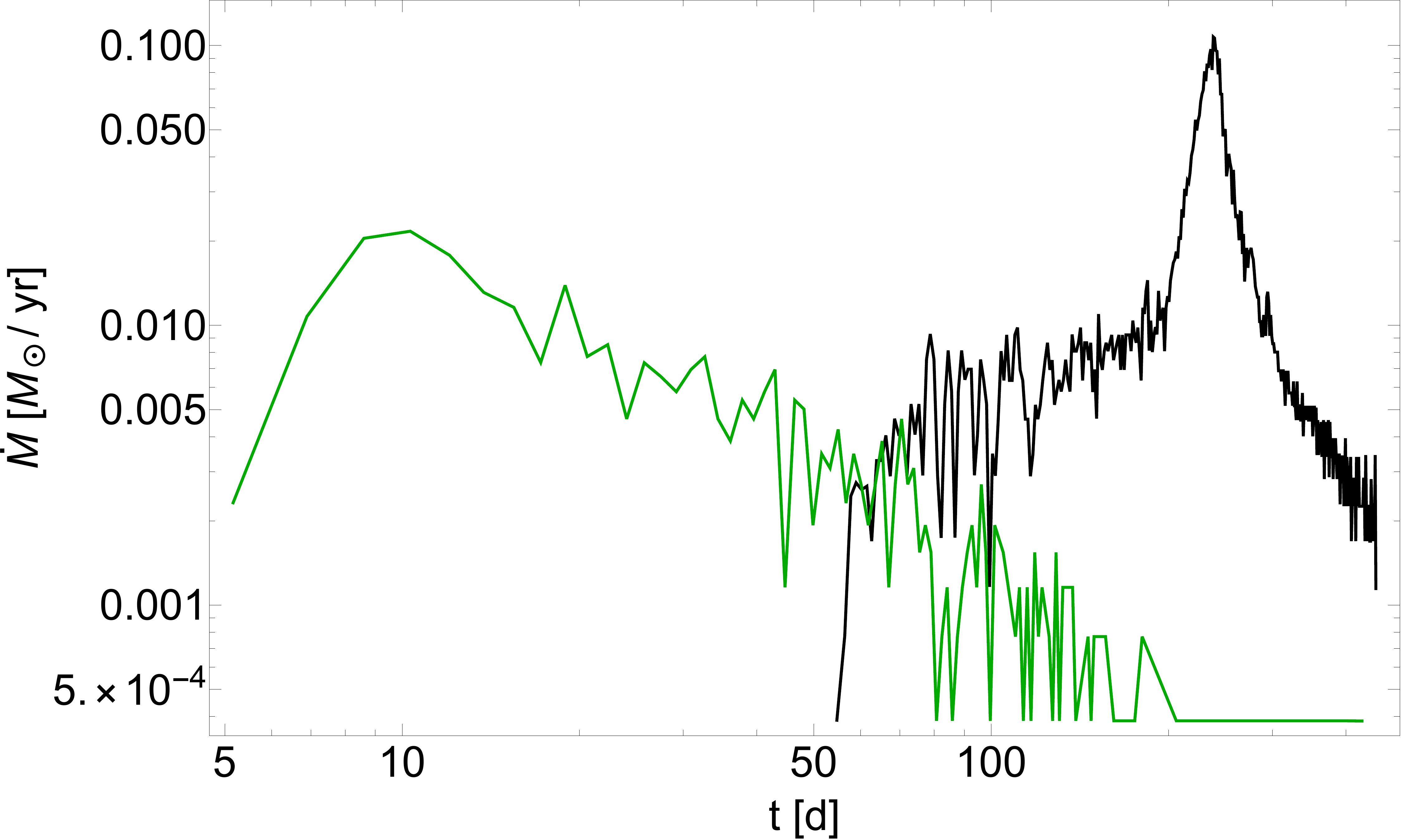} 
   \includegraphics[width=0.495\textwidth]{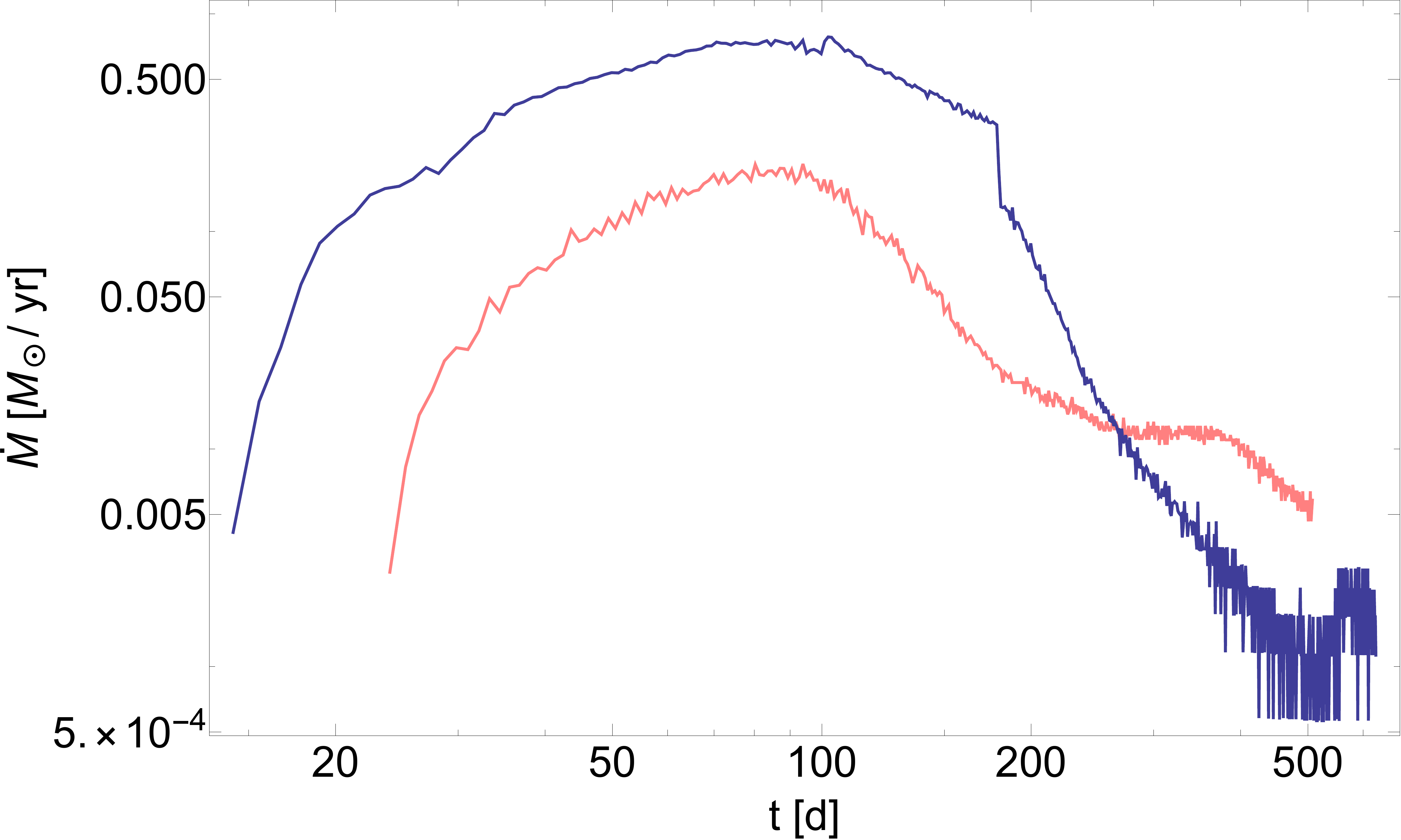} 
   \caption{{}{The top-left, top-right, and middle-left plots show }the accretion rate in Solar masses per year as a function of time in days for $q = 0.005$ and $a = 2.5$ mpc, which is the setup most favored by the scaling arguments for the event ASASSN-15lh. The dashed curve in the top-left panel shows the analytic prediction that follows from the impulse approximation, and the horizontal, dot-dashed line shows the Eddington limit assuming $L = 0.1\dot{M}c^2$. The different curves in each panel -- aside from the analytic solution -- are from different simulations, and are only plotted separately for the sake of clarity. {}{The middle-right, bottom-left, and bottom-right panels illustrate} the accretion rate in Solar masses per year as a function of time in days for $q = 0.005$ and $a = 5$ mpc. }
   \label{fig:mdots_qp005_a500}
\end{figure*}

\subsection{Results}
Of the 30 disruptions with $a = 2.5$ mpc and $q = 0.005$, 23 generated a completely unbound stream, thereby resulting in no accretion, or a stream on a wide orbit about the binary {}{that, in some instances, encircled the system in a circumbinary ring of gas that eventually spread into a disc \citep{coughlin17b}. These 23 cases had little to no} accretion at all times, bearing little resemblance to a ``normal'' TDE.

{}{As pointed out in \citet{coughlin17}, the reason for these ejections and atypical stream distributions is that the specific energy of the center of mass of the star is modified by the motion of the secondary. In particular, in the rest frame of the secondary, the star has an approximate energy of $\epsilon \simeq \mathbf{v}_{app}^2/2-GM/r$, where $\mathbf{v}_{app}^2$ is the square of the vector difference between the velocity of the star and the velocity of the secondary. In general, $\mathbf{v}_{app}^2$ contains cross terms that depend on the relative orientation of the star and the secondary at the time of disruption, and will be smaller (larger) if the star and the secondary are moving parallel (antiparallel) to one another when the star enters within the tidal radius. However, these cross terms sum to approximately zero when averaged over the ensemble of disruptions, and the net contribution to the average energy of the star is given by the kinetic energy of the secondary. In the small-$q$ limit, $\epsilon = GM_p/(2a)$, $M_p$ being the mass of the primary, and the ratio of this energy to the tidal energy spread $\Delta \epsilon$ is $\epsilon/\Delta\epsilon \simeq q^{-1/3}(M_p/M_*)^{1/3}/(2k)$, where $k$ is the ratio of the binary separation to the tidal radius of the primary (see also Equation 8 of \citealt{coughlin17}). For the present setup, this ratio is $\epsilon/\Delta\epsilon \simeq 6$ when $a = 2.5$ mpc ($\simeq 3$ when $a = 5$ mpc), meaning that the orientation of the disrupted star must be favorable in order to cancel the relatively large contribution to the energy that results from the motion of the secondary. Instances in which the alignment is not favorable result in a completely unbound stream, and it is therefore not surprising that the majority of the simulated disruptions generate little to no accretion.} 

The other 7 yielded accretion at early times, with some likelihood of appearing like a TDE from an isolated SMBH, and {}{the top-left, top-right, and middle-left panels of} Figure \ref{fig:mdots_qp005_a500} illustrate the accretion rate onto the secondary SMBH that results from these disruptions. Each curve represents a different simulation, except for the dashed curve in the top-left hand panel, which shows the solution that follows from the impulse approximation for comparison, and the horizontal, dot-dashed line in the same panel, which gives the Eddington limit of the hole assuming $L = 0.1\dot{M}c^2$. The accretion curves were broken up into separate panels for clarity.

It is obvious that the accretion curves do not match the impulse approximation exactly, and there are a number of reasons for this discrepancy; for one, the impulse approximation measures the ``fallback rate,'' being the rate at which material returns to pericenter. On the other hand, material that accretes onto the black hole in the simulation must first transfer its angular momentum outwards and dissipate its kinetic energy. It is also apparent that the return time of the most bound debris is either earlier or later than that predicted analytically, and this results not only from the fact that the star is already tidally distorted at the time it passes through the tidal radius \citep{coughlin15}, but also because the binding energy of the center of mass is not exactly equal to zero. In particular, the three-body interactions change the specific energy of the star from its original, parabolic value, meaning that the most bound debris has an energy that deviates from the value imposed by the tidal force and thus returns at an earlier or later time. 

For the same reason, the peak accretion rate can differ from the analytically-predicted one by over an order of magnitude, and the total mass accreted can exceed or fall below the expected value of $\simeq M_\odot/2$. Nevertheless, the time to peak measured from the start of accretion is typically around the same value, and corresponds to roughly the value that results from the impulse approximation ($\sim 30-40$ d). Consistent with this observation, one can show, by extending the analysis in \citet{lodato09} and \citet{coughlin14} to include a non-zero center of mass energy $\epsilon_c$, that the time to rise is generally insensitive to $\epsilon_c$ as long as $\epsilon_c \lesssim \Delta\epsilon$, where $\Delta\epsilon = GM_hR_*/r_t^2$. 

The presence of the binary companion also generates a number of perturbations on the accretion rate, including, as was noted in \citet{coughlin17}, sudden and drastic increases and decreases in its magnitude. We also see from each of the three panels that there is a time at which the accretion rate first exhibits a large change, and this corresponds roughly to a time of 100-150 days following the disruption of the star. We will return to a more in depth discussion of this point and its implications for ASASSN-15lh in Sections \ref{sec:accretion} and \ref{sec:asassn15lh}.

The {}{middle-right, bottom-left, and bottom-right panels} of Figure \ref{fig:mdots_qp005_a500} show the accretion rates from disruptions by a binary with $q = 0.005$ and $a = 5$ mpc; as was true for those with $q = 0.005$ and $a = 2.5$ mpc, the six shown in this figure looked like standard disruptions and had prompt accretion, while the other 14 resulted in completely ejected streams or low-amplitude, variable, and late-time accretion. The main features of these plots are similar to those with $a = 5$ mpc, but we note that the time at which the accretion rate starts to show significant variability is typically closer to 200 days. 

\begin{figure*} 
   \centering
   \includegraphics[width=0.495\textwidth]{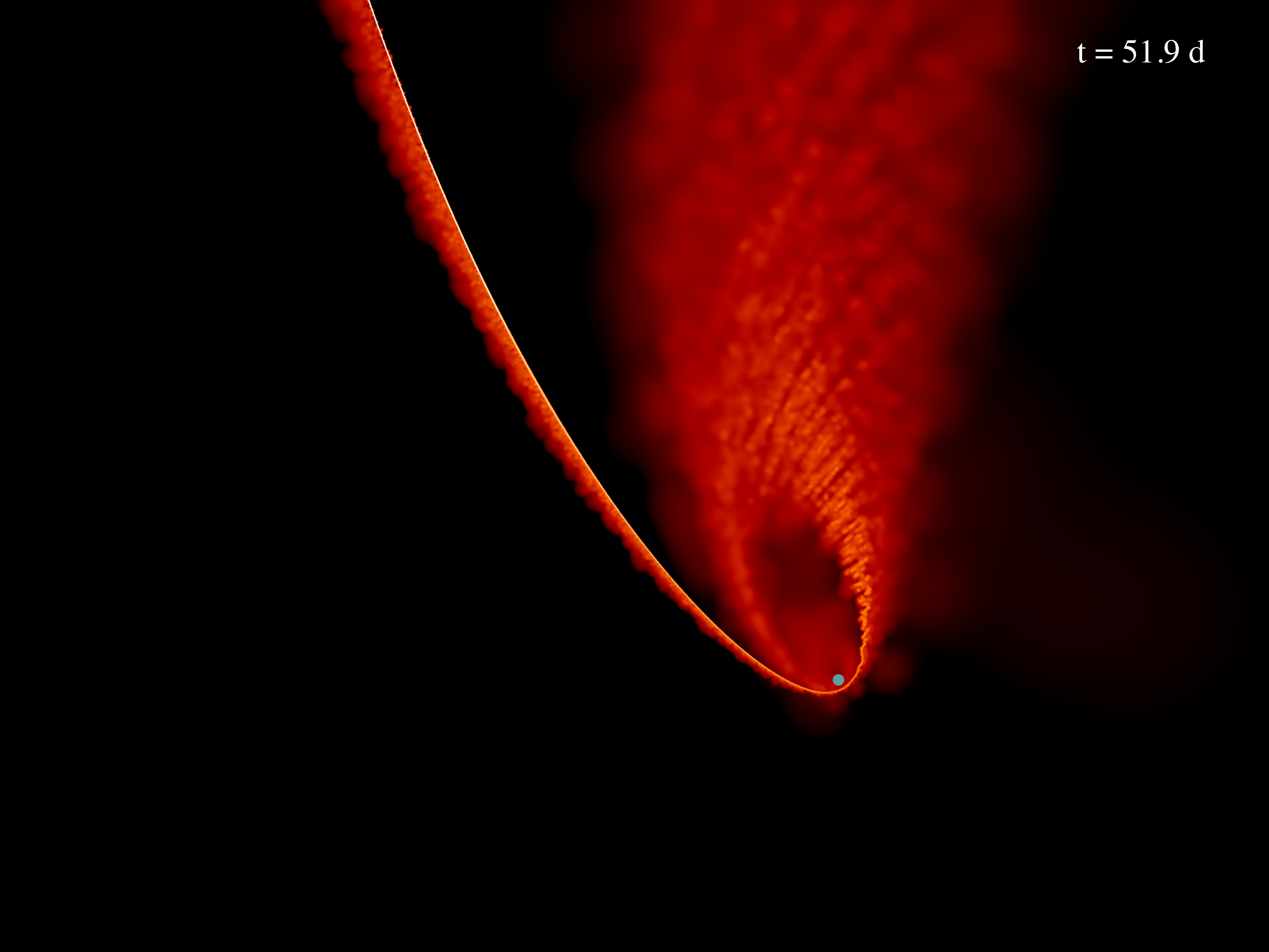} 
   \includegraphics[width=0.495\textwidth]{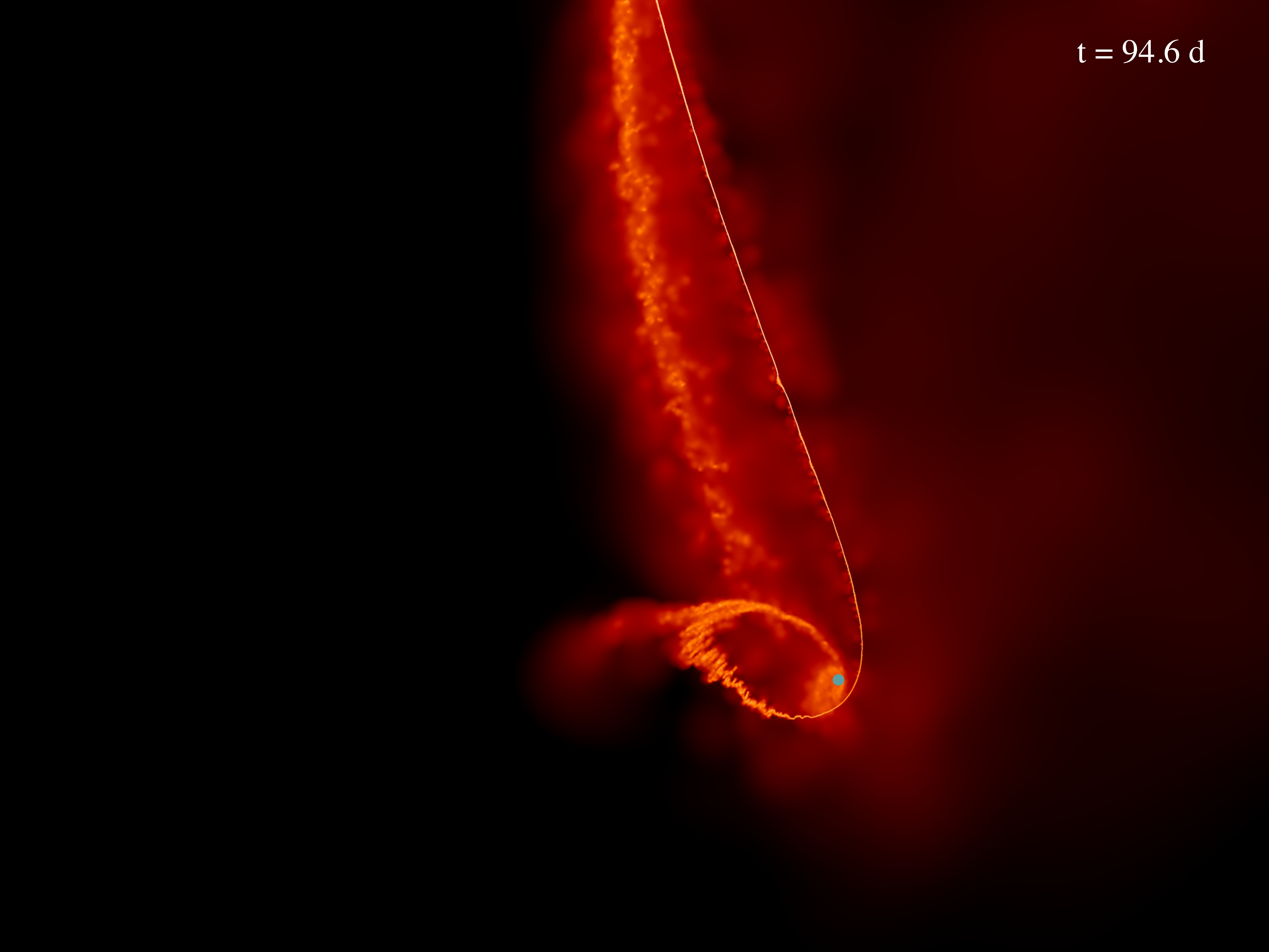} 
   \includegraphics[width=0.495\textwidth]{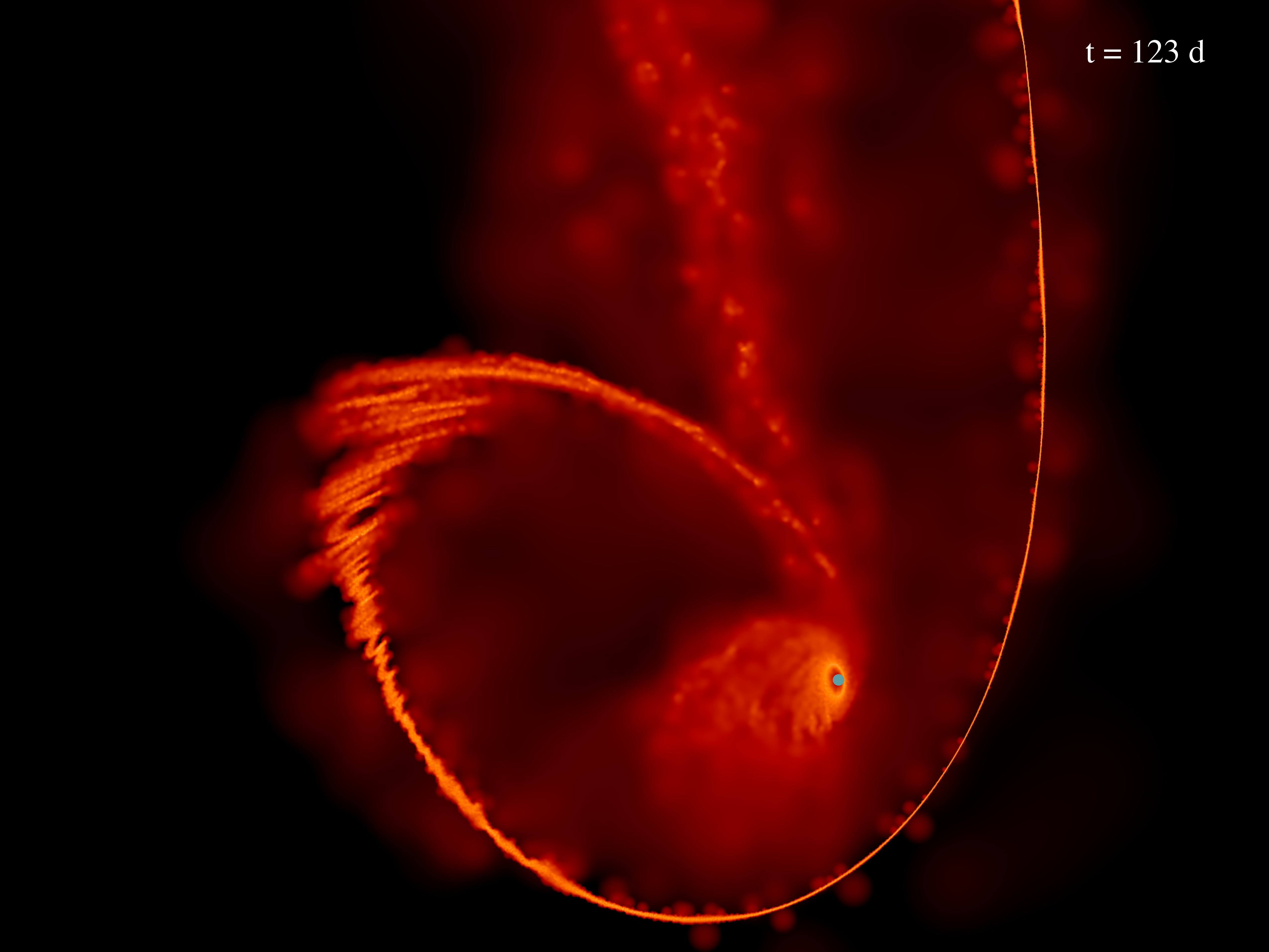} 
      \includegraphics[width=0.495\textwidth]{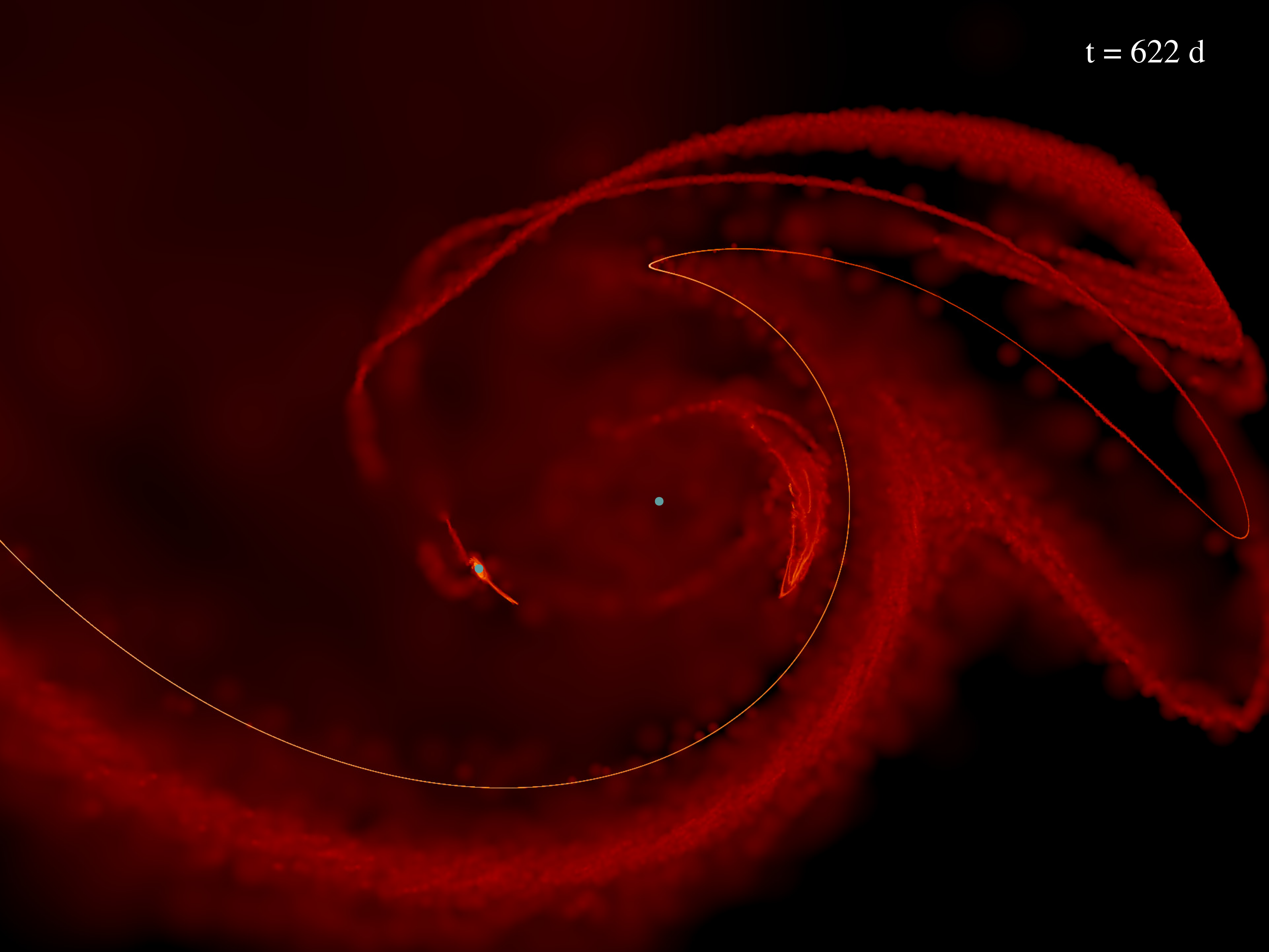} 
   \caption{{}{The top-left, top-right, and bottom-left show an} example of the accretion disk that forms around the secondary SMBH from a TDE. This simulation corresponds to the accretion rate shown by the blue curve in the middle-left panel of Figure \ref{fig:mdots_qp005_a500}, and the time since disruption is shown in the top-right corner of each panel. Color scales with density, and brighter (darker) colors indicate denser (less dense) regions. {}{The bottom-right panel of this figure shows the same simulation at a later time, and expanded in scale to demonstrate that, in addition to maintaining a small-scale accretion disc around the secondary, some debris expands to large radii and surrounds the binary. The blue points show the positions of the SMBHs.} }
   \label{fig:disks}
\end{figure*}

The morphology of the accretion flows that result from these simulations are similar in appearance to those in \citet{coughlin17}, with a small-scale accretion disk around the secondary (disrupting) SMBH and a larger-scale, extended cloud of debris that surrounds the binary; see Figure \ref{fig:disks}. However, while accretion onto both black holes occurred most of the time for the moderate-$q$ binaries considered in \citet{coughlin17} (see their Figure 12 and the online, supplementary material), the simulations here generally did not result in accretion onto the primary (or, if it did, it was at late times and was very low). This could be due either to the difference in mass ratio or to the wider separations considered here -- \citet{coughlin17} let $a = 100\,r_{t,1}$, while here $a = 500$ -- $1000\,r_{t,1}$. 

\section{Discussion}
\label{sec:discussion}
\subsection{Accretion rates and morphology}
\label{sec:accretion}
It is apparent from Figure \ref{fig:mdots_qp005_a500} that, for the simulations that result in prompt accretion onto the secondary black hole, there is an initial rise, peak, and decay in the accretion rate that resembles the isolated-SMBH, canonical-TDE picture. After a finite amount of time, however, the accretion rate changes drastically, and can decrease (e.g., the green curve in the top-left panel of Figure \ref{fig:mdots_qp005_a500}) or increase (e.g., the blue curve in the middle-left panel of Figure \ref{fig:mdots_qp005_a500}) by an order of magnitude and do so rather abruptly. 

We argued in Section \ref{sec:timescales} that, when a TDE occurs by the secondary SMBH, such deviations are expected to occur in systems with small $q$, the reason being that the accreting debris recedes outside the Hill sphere of the secondary and affects the accretion rate on a timescale given by Equation \eqref{tsig}. Using $a = 2.5$ mpc and $M_1 = 10^8M_{\odot}$ in this expression gives $T_\sigma \simeq 90$ days, which is in rough agreement with Figure \ref{fig:mdots_qp005_a500}. We also find $T_{\sigma} \simeq 250$ days for $a = 5$ mpc and $M_1 = 10^8M_{\odot}$, which is approximately the timescale at which changes start to occur in the accretion curves shown in the {}{middle-right, bottom-left, and bottom-right panels }of Figure \ref{fig:mdots_qp005_a500}.

The relative orientation of the returning debris stream, the secondary black hole, and the center of mass of the binary -- approximately at the location of the primary -- determine the magnitude of the change in the accretion rate. Furthermore, if this interpretation is correct, the orientation and physical extent of the accretion disk around the secondary should reflect these changes, as the angular momentum profile of the incoming debris, relative to the secondary, starts to differ from the material initially comprising the disk.

In support of this notion, {}{the top-left, top-right, and bottom-left panels of }Figure \ref{fig:disks} show the accretion disc around the secondary SMBH at three different times, those times given in the top-right corner of each panel; in this figure, colors scale with the density, with brighter (darker) colors representing denser (more rarefied) material. This simulation corresponds to the accretion rate shown by the blue curve in the middle-left panel of Figure \ref{fig:mdots_qp005_a500}. We see that initially a large-scale, elliptical disk forms around the SMBH, and the accretion rate starts to rise and resemble the situation from an isolated TDE. However, after $\sim 90$ days, the orientation of the debris stream starts to change, and material is funneled into a much closer region around the black hole. After this time, a new, much more compact disk forms around the SMBH, and its rotation profile is retrograde with respect to the accretion disk that originally formed. In this case, the debris stream, the secondary, and the center of mass are nearly aligned, and this alignment results in a decrease in the angular momentum of the incoming material and a corresponding increase in the accretion rate by over an order of magnitude. The bottom-right panel shows a zoomed-out version of the same simulation at a later time, demonstrating that some debris is flung to larger distance and generates a cloud of debris that encompasses the binary. 

In addition to an initial, major increase or decrease in the accretion rate, many of the simulations exhibit variability at later times. This behavior was also seen in the simulations run by \citet{coughlin17}, and these additional variations could occur when the secondary SMBH comes close to the debris stream; while \citet{liu09} and \citet{ricarte16} argued that these instances should cause dips in the accretion rate of the primary, here we see -- not surprisingly -- that they can also result in an increase in the accretion rate of the secondary. 


\begin{figure}
\centering
\includegraphics[width=0.49\textwidth]{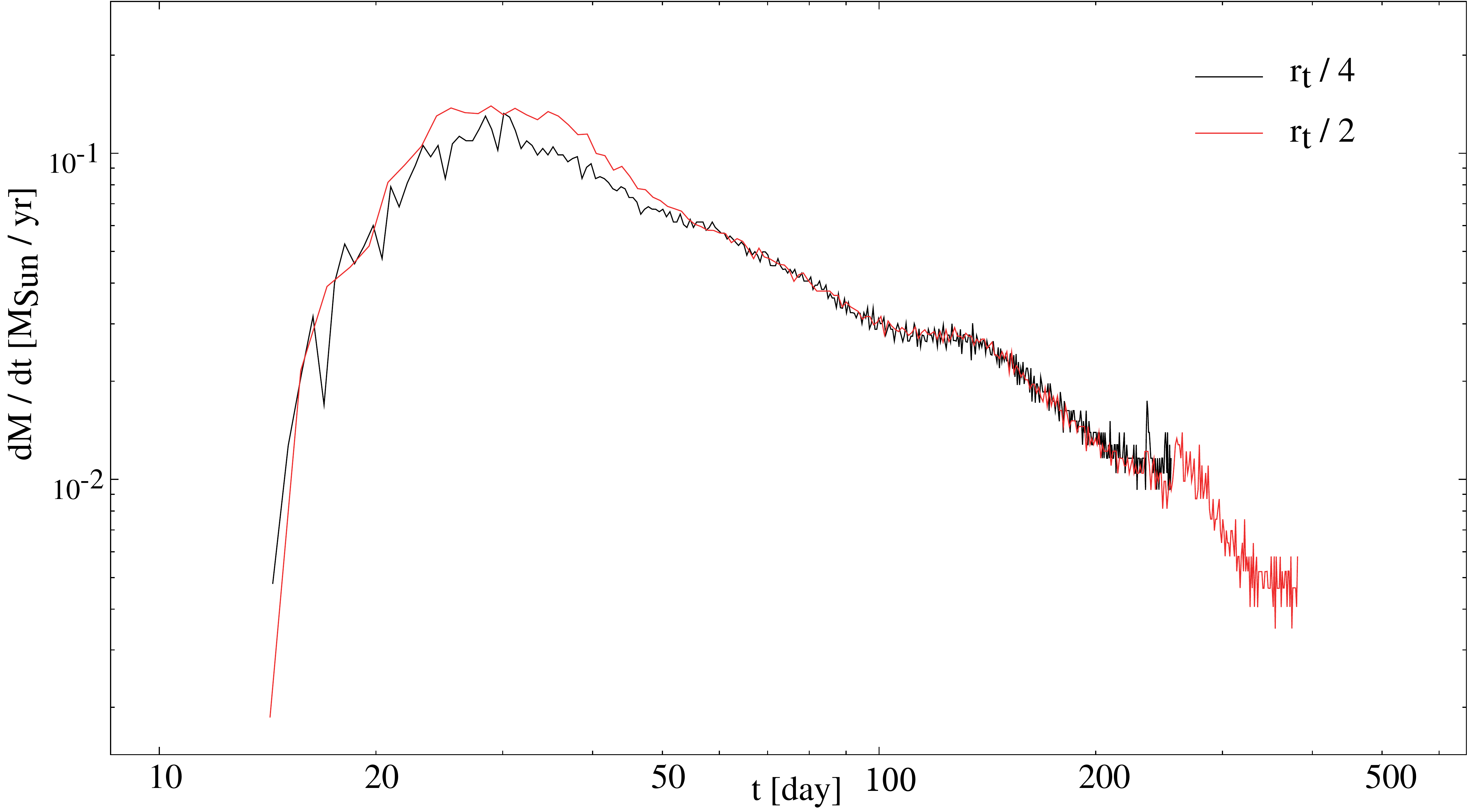}
\caption{The accretion rate onto the SMBH for two values of the accretion radius, as shown in the legend. The red curve in this figure is the same as the yellow curve in the middle-left panel of Figure \ref{fig:mdots_qp005_a500}.}
\label{fig:mdots_comps}
\end{figure}

Some of the changes in the accretion rates in Figure \ref{fig:mdots_qp005_a500} are likely due to the specific choice of accretion radius; for example, the near-vertical spikes occur when the stream passes into or out of the accretion radius of the hole. However, these changes are ultimately caused by the change in the orientation of the stream, as evidenced by Figure \ref{fig:disks}, and choosing a smaller accretion radius would likely serve to reflect the same features but smoothed out temporally. Additionally, we speculate that the more gradual changes in the accretion curves are less sensitive to the accretion radius, and Figure \ref{fig:mdots_comps} confirms this notion: the red curve shows the accretion rate when the accretion radius is set to $r_t/2$ (and is identical to the yellow curve shown in the middle-left panel of Figure \ref{fig:mdots_qp005_a500}), and the black curve illustrates the accretion rate when the accretion radius is set to half that value. It is apparent that, for this TDE, only small changes in the accretion rate are induced by adopting a smaller accretion radius.

{}{Before moving on, we note that the relatively modest particle number ($5\times 10^5$ particles---chosen because higher resolution would be substantially more expensive) results in numerical dissipation in the immediate vicinity of the accreting SMBH. This likely leads to faster stream circularization and disc formation than would be the case if there were only physical sources of dissipation. In the physical situation, apsidal precession due to general relativity is responsible for stream self-intersection and circularization, and without this effect there would be a significant delay in the onset of accretion \citep{bonnerot16,hayasaki16,shiokawa15}. In the case of close binaries, the binary potential will cause additional precession that increases with distance from the disrupting black hole. Circularization will therefore occur on a time scale that is no longer (and possibly shorter) than for the case of an isolated black hole. Thus, while disc formation may be artificially enhanced by numerical effects in our models, there are physical sources of precession that correspondingly generate physical dissipation on timescales comparable to the fallback time.}

\subsection{Application to ASASSN-15lh}
\label{sec:asassn15lh}
ASASSN-15lh was among the brightest supernovae ever detected, with an estimated peak bolometric luminosity of $L \simeq 10^{45}$ erg s$^{-1}$ and a total radiated energy in excess of $10^{52}$ erg \citep{dong16,godoy17}. In addition, the event rebrightened -- only in the UV bands -- around day $\sim 120$ after first detection, reached a second peak around day $\sim 200$, and then continued to decline \citep{brown16,godoy17}. As noted by \citep{margutti17}, who found that the source was emitting in the soft X-ray in addition to the optical and UV, ASASSN-15lh also exhibited a significant degree of variability in the UV bands during the rebrightening phase; there may have also been a second rebrightening, only in the shortest wavelengths, around 260 days post-detection \citep{brown16}.

\citet{dong16} pointed out that the lack of any hydrogen emission lines means that the rebrightening is likely unassociated with the interaction between the ejecta from a supernova and circumstellar material (but see \citealt{chatzopoulos16}); timescale arguments also suggest that this association is unlikely \citep{margutti17}. Likewise, the peak luminosity of ASASSN-15lh would require a very large amount ($\gtrsim 30M_{\odot}$) of $^{56}$Ni, and the decline in luminosity is inconsistent with the radioactive decay timescale \citep{dong16,quimby11}. While a magnetar-powered explosion is still possible, there is only a narrow range of permissible properties of the neutron star if one is to accommodate all aspects of the observations \citep{chatzopoulos16}. Finally, the properties of the host galaxy, being very massive with little star formation, are inconsistent with those of other, superluminous supernovae \citep{neill11}.

While the proximity of the event with the nucleus of the host galaxy \citep{brown16} favors the tidal disruption scenario for the origin of ASASSN-15lh (the lack of hydrogen emission lines is also consistent, depending on the orientation of the disrupted debris; \citealt{guillochon14,roth16}), arguably the biggest criticism of this model is that the mass-luminosity relation implies a SMBH mass of $\sim 10^{8}M_{\odot}$ \citep{mcconnell13}. A straightforward calculation shows that SMBHs of such mass (and larger) consume-Solar type stars (and smaller) whole. Nevertheless, the tidal radius can be pushed slightly outside the event horizon if the hole is rapidly rotating in a prograde sense with respect to the angular momentum vector of the stellar orbit\citep{kesden12}, and such a scenario was invoked by \citet{leloudas16} to explain ASASSN-15lh. 

On the other hand, if one posits that a lower-mass, companion SMBH was the disrupting hole, then such a black hole can have a mass well below the maximum-allowable mass for disruption. As we showed above, one can use the rise time, which was successfully inferred for this event, to constrain the mass of the secondary to $M_h \simeq 5\times10^{5}M_{\odot}$, and the time at which rebrightening occurs gives $a \simeq 3.6$ mpc. 

Our hydrodynamic simulations with $a = 2.5$ mpc, $q = 0.005$, and $M_1 = 10^8M_{\odot}$ generated a number of interesting fallback signatures, and many of them do not look like ASASSN-15lh. A few, however, do bear a very similar resemblance to the event. For example, the yellow accretion curve middle-left panel of Figure \ref{fig:mdots_qp005_a500} (see also Figure \ref{fig:mdots_comps}) has a rise time on the order of $\sim 30$ days, a power-law decay following the initial peak, and a second rebrightening that starts around $\sim 100$ days. This rebrightening event lasts for roughly 60 days, after which the accretion rate resumes its power-law decline until much later times. The purple curve in the top-left panel of Figure \ref{fig:mdots_qp005_a500} also has many of the same features, with an initial peak after $\sim 60$ days, a second brightening starting at $\sim 150$ days that lasts $\sim 100$ days, and otherwise an approximate power-law decline (again, until much later times). 

As can be seen from Figure \ref{fig:disks}, the increase in the accretion rate corresponds to a substantial restructuring of the accretion disk around the hole, with more material being funneled to small radii. If the disk remains thin, then a natural consequence of such a small-scale accretion disk would not only be an increase in the accretion rate, but also an increase in the effective temperature \citep{shakura73}. Therefore, this model immediately incorporates the larger UV flux seen from ASASSN-15lh. 

The peak luminosity of the event reached $L\simeq 10^{45}$ erg s$^{-1}$, which, for a SMBH mass of $5\times10^{5}M_{\odot}$ and a radiative efficiency of $0.1$, corresponds to a few times the Eddington limit of the hole. Accretion rates of this magnitude are achieved by our simulations, and the fact that the luminosity is only mildly supercritical is consistent with the lack of hard X-ray emission that would be expected from highly super-Eddington TDEs (such as, for example, \emph{Swift} J1644+57; e.g., \citealt{bloom11, burrows11, cannizzo11,levan11,zauderer11}). Furthermore, the total radiated energy, $E \simeq few\times10^{52}$ erg, is manifestly obtained if the disrupted star is Solar-like and half of the star is accreted (again, with a radiative efficiency of 0.1). While the shift in center of mass energy means that some of our simulations accrete more or less of half of a Solar mass, the scale of the total radiated energy is always of the order $few \times10^{52}$ erg.

Finally, many of the curves in Figure \ref{fig:mdots_qp005_a500} show some degree of variability in the accretion rate near the time at which the rebrightening occurs. While the magnitude of the variations are certainly somewhat dependent on the size of the accretion radius and the resolution of the simulation, since the change in accretion morphology during the rebrightening is rather drastic, we expect some intrinsic, physical variability around this time, which is consistent with that seen in ASASSN-15lh \citep{margutti17}. Additionally, most of the accretion rates show some other resurgence over the roughly power-law decline at even later times (i.e., after the rebrightening that we expect to occur on the timescale given by Equation \ref{tsig}). As noted by \citet{brown16}, there is some evidence that ASASSN-15lh underwent a second rebrightening around day $\sim 260$, but only in the highest-energy bands. Not only are such additional variations in line with the binary SMBH interpretation, we might also expect -- if the binary SMBH interpretation is correct -- the event to exhibit some fluctuations in its lightcurve at later times still. 

\subsection{Rate of TDEs}
In our model, the star is tidally disrupted by the secondary, which has a mass $M_2 = 0.005M_1$. As shown by \citet{coughlin17}, who performed a more systematic study of the effects of $q$ on the properties of TDEs by SMBHs in binaries, the relative probability of disruption by the secondary SMBH is approximately $\lambda_2 \simeq q/2$. Thus, the vast majority of stars in systems with $q = 0.005$ should be disrupted by the primary. 

In the setup considered here, where the primary mass is $10^8M_{\odot}$, all of the stars that are ``disrupted'' by the primary are actually swallowed whole, as $r_{g,1} = GM_1/c^2 \simeq r_{t,1}$. Thus, while it would seem unlikely that any observed TDE in such a small-$q$ system should be committed by the secondary, we circumvent this issue -- stars are much more likely to enter within the tidal radius of the primary, but the act of doing so generates no observable emission. 

Nevertheless, the total rate of tidal disruption for the systems considered here is rather low: of a total of $6\times10^6$ encounters with $a = 2.5$ mpc, the restricted three-body integrations resulted in only 44 disruptions by the secondary (and we simulated the hydrodynamic evolution for 30 of those). Using the fact that the separation of the binary is $500r_{t,1}$, the rate of disruption by the binary is $\lambda_{TDE,binary} \simeq 44\times500/(6\times10^6)\lambda_{TDE} \simeq 0.004\lambda_{TDE}$, where $\lambda_{TDE}$ is the rate of disruption by an isolated SMBH. {}{Moreover, if we only consider the seven disruptions that generated prompt accretion, then the \emph{observable} rate of disruption falls by an additional factor of $\sim$ 7. }If $\lambda_{TDE} = 10^{-4}$ gal$^{-1}$ yr$^{-1}$ \citep{frank76,stone16}, then we expect roughly 0.2 TDEs over the lifetime of the binary{}{, and even fewer that result in observable accretion}. When $a = 5$ mpc, the total number of expected TDEs increases to $\sim 3$ before the binary inspirals due to gravitational wave losses. Thus, the likelihood of tidally disrupting a Solar-like star is somewhat low over the lifetime of the binary.

On the one hand, superluminous supernovae are quite rare, and ASASSN-15lh is among the most extraordinary within that class. It would therefore not be overly surprising if the event itself also required a set of extraordinary conditions. Given the large number of relatively low-mass satellite galaxies, it also seems plausible that there may exist a significant population of low-mass black holes spiraling into the centers of larger galaxies. 

Another possibility is that our initial conditions for the three-body encounters -- a star approaching the binary center of mass on a parabolic trajectory -- were too restrictive. As shown by \citet{chen09}, if there is a significant population of bound stars to the primary, this can boost the TDE rate by a factor of $\sim 10^4$. Stars bound to the secondary could also undergo Kozai oscillations, and those with large orbital inclinations could be driven to such high eccentricies (the eccentrici Kozai mechanism) that they plunge within the tidal radius of the secondary \citep{ivanov05,li15}; this could also significantly augment the TDE rate. Finally, as shown by \citet{madigan17}, an eccentric disk of stars maintains its stability by pushing some stellar orbits to high eccentricities, which -- if such a population of stars is present, and some studies suggest they may be somewhat common \citep{lauer05} -- further increases the TDE rate.

\subsection{Other Sources of Emission: Winds and Accretion onto the Primary}
From Figure \ref{fig:mdots_qp005_a500}, we see that accretion onto the secondary often proceeds at a super-Eddington rate. While most of the time the rate is only mildly super-Eddington, satisfying $\dot{M} = N\times\dot{M}_{Edd}$ with $N \simeq few$, in some cases, and particularly toward the peak in the accretion rate, $N$ can approach and even exceed 100. While these hyper-Eddington periods are predicted analytically \citep{rees88,evans89}, the peak accretion rates in Figure \ref{fig:mdots_qp005_a500} can actually exceed the analytical one (shown by the dashed curve in the top-left panel in this figure), which is a consequence of self-gravity \citep{coughlin15} and the non-zero energy of the stellar center of mass.

In these highly supercritical regimes, one might expect winds \citep{strubbe09,metzger16} or, in the more extreme cases, jets \citep{giannios11,coughlin14,tchekhovskoy14,coughlin15b} to be launched from the vicinity of the accreting SMBH. From these outflows one would expect an additional source of radiation, with properties distinct from the disc. While it is unlikely that this scenario explains the rebrightening seen in ASASSN-15lh -- the accretion rate would have been super-Eddington prior to the increase, and the constancy of the X-ray emission \citep{margutti17} suggests that there was no change in the source of harder photons -- changes in the luminosity induced by a sudden super-Eddington surge of accretion would be possible in other systems. 

The disk feeding the SMBH should also become geometrically thicker once the Eddington limit is surpassed, and could perhaps inflate to a quasi-spherical envelope once $L \gg L_{Edd}$ \citep{loeb97,coughlin14}. The simulations performed here would not capture the important physical interactions between the radiation and the plasma that would lead to such an inflated disk structure (see \citealt{sadowski16}, who do find quasi-spherical envelopes from their simulations of TDEs). Nevertheless, if the photosphere of the disk coincides with the photon trapping radius \citep{begelman79}, then we can estimate the size of the envelope as (Equation 17 of \citealt{coughlin14})

\begin{multline}
R_{env} \simeq \left(\frac{\kappa}{4\pi{c}}M_*\sqrt{GM_h}\right)^{2/5} \\ \simeq 7\times10^{14}\left(\frac{M_*}{M_{\odot}}\right)^{2/5}\left(\frac{M_h}{5\times10^{5}M_{\odot}}\right)^{1/5} \text{ cm},
\end{multline}
where in the final line we set the opacity equal to the value for electron scattering, $\kappa = \kappa_{\rm es} = 0.34$ g cm$^{-2}$. Interestingly, this radius is on the order of the Hill sphere of the secondary for the binaries considered here; thus, the disks from super-Eddington TDEs in tight binaries could fill their Roche Lobes, resulting in accretion onto the primary. This accretion would then provide an additional source of radiation, and it would also modify the structure of the envelope around the secondary and, conceivably, its accretion rate. 

While it did not occur in any of our simulations, one can also imagine situations in which the unbound debris from the initial disruption is captured by, and accretes onto, the primary. Using the fact that the terminal velocity of the unbound debris is $v_{\infty} = v_{\rm esc}(M_h/M_*)^{1/6}$ \citep{rees88}, where $v_{\rm esc} = \sqrt{2GM_*/R_*}$ is the escape velocity of the star, one would expect accretion onto the primary to begin on a timescale of

\begin{equation}
T_{acc,1} \simeq \frac{a}{v_{\infty}} \simeq 1.1\left(\frac{a}{2.5\text{ mpc}}\right)\left(\frac{v_{\text{esc}}}{v_{\text{esc},\odot}}\right)^{-1}\left(\frac{M_h}{5\times10^{5}M_{\odot}}\right)^{-1}\text{yr}.
\end{equation}
This number is really a lower limit, as the terminal velocity of the unbound debris is not reached instantaneously. The debris is also likely to have some angular momentum with respect to the primary, and so accretion will not begin immediately. Indeed, this situation is encountered in some of our simulations, and accretion onto the primary is very slow (and virtually nonexistent). 

\section{Summary and Conclusions}
\label{sec:summary}
In this paper we analyzed the tidal disruption of Solar-like stars by the secondary black hole in binary systems with high mass primaries and extreme mass ratios ($q = M_2/M_1 = 0.005$). These systems are interesting because (1) for sufficiently high primary masses no TDEs from the primary are expected, and (2) for extreme mass ratios the binary can be tight enough to perturb a secondary TDE while the gravitational wave inspiral time is still moderately long (1-10~Myr for our specific parameters). We argued heuristically that these events should display lightcurves that initially rise, peak, and decay in a manner similar to those of TDEs by isolated SMBHs. However, after a time $T_{\sigma} \propto {a^{3/2}}/\sqrt{GM_1}$, where $a$ is the separation of the binary and $M_1$ is the mass of the primary, the accretion onto the secondary should exhibit some deviation from the canonical-TDE scenario, resulting in a variation in the lightcurve.

Using a combination of 3-body integrations and hydrodynamic simulations, we studied the properties of the fallback from TDEs by the secondary in systems with black hole masses of $10^8 \ M_\odot$ and $5 \times 10^5 \ M_\odot$, and $a=2.5 \ {\rm mpc}$ or 5~mpc. The most common outcome---occuring in about 2/3 to 3/4 of realizations---was the formation of an unbound debris stream that would not result in any accretion (though it may mimic a supernova as it slams into the circumnuclear environment; \citealt{guillochon16}), or a very weakly-bound stream with accretion at late times that would not yield a recognizable TDE signal. The subset of bound debris events, however, produced simulated accretion curves that displayed structure on approximately the analytically-predicted timescales. As in the case of more nearly equal mass binaries, studied by \citet{coughlin17}, a variety of morphologies were produced, including cases where the accretion rate showed sudden declines or excesses at late times. We associate these features with binary-induced changes in the specific angular momentum of the fallback material.

Based on our results, we suggest that two classes of SMBH binaries are potentially detectable from their effect on TDE lightcurves. The first is systems with low mass primaries ($M_1 \sim 10^6 \ M_\odot$) and moderate $q$, characteristic of sub-Milky Way mass galaxies. Disruptions by either the primary or the secondary might occur in these systems, though the rate of primary disruptions dominates. The second is systems with primary masses above the mass where Solar-type stars are swallowed whole (about $10^8 \ M_\odot$, though with some dependence on the spin). There could be a significant population of extreme mass ratio secondaries in such systems, and these secondaries could generate TDEs with distinctive ``binary" lightcurves at separations of the order of 1-10~mpc. The rate is uncertain, though probably low, but such events might be identifiable given a large enough sample of massive host galaxies where primary TDEs are not possible.

Finally we applied our results to the puzzling transient ASASSN-15lh , which has variously been identified as an exceptionally luminous SN or as a TDE. If it is a TDE, the host galaxy properties imply a primary mass that is close to or above the critical value of $10^8 \ M_\odot$, making the alternate hypothesis of a secondary disruption conceivable. We make no strong claims, but note that some of the features of the ASASSN-15lh event---most particularly its rebrightening about 100 days after detection---are reasonably common outcomes of our simulations. ASASSN-15lh could be an example of a TDE by a low mass secondary in a tight binary with a massive primary.

\section*{Acknowledgments}
Support for this work was provided by NASA through the Einstein Fellowship Program, grant PF6-170150. PJA acknowledges support from NASA through grant NNX16AI40G. {}{We thank the referee for useful comments and suggestions.} We used {\sc splash} \citep{price07} for the visualization. This research used the Savio computational cluster resource provided by the Berkeley Research Computing program at the University of California, Berkeley (supported by the UC Berkeley Chancellor, Vice Chancellor for Research, and Chief Information Officer).

\bibliographystyle{mnras}
\bibliography{refs}

\begin{thebibliography}{}
\makeatletter
\relax
\def\mn@urlcharsother{\let\do\@makeother \do\$\do\&\do\#\do\^\do\_\do\%\do\~}
\def\mn@doi{\begingroup\mn@urlcharsother \@ifnextchar [ {\mn@doi@}
  {\mn@doi@[]}}
\def\mn@doi@[#1]#2{\def\@tempa{#1}\ifx\@tempa\@empty \href
  {http://dx.doi.org/#2} {doi:#2}\else \href {http://dx.doi.org/#2} {#1}\fi
  \endgroup}
\def\mn@eprint#1#2{\mn@eprint@#1:#2::\@nil}
\def\mn@eprint@arXiv#1{\href {http://arxiv.org/abs/#1} {{\tt arXiv:#1}}}
\def\mn@eprint@dblp#1{\href {http://dblp.uni-trier.de/rec/bibtex/#1.xml}
  {dblp:#1}}
\def\mn@eprint@#1:#2:#3:#4\@nil{\def\@tempa {#1}\def\@tempb {#2}\def\@tempc
  {#3}\ifx \@tempc \@empty \let \@tempc \@tempb \let \@tempb \@tempa \fi \ifx
  \@tempb \@empty \def\@tempb {arXiv}\fi \@ifundefined
  {mn@eprint@\@tempb}{\@tempb:\@tempc}{\expandafter \expandafter \csname
  mn@eprint@\@tempb\endcsname \expandafter{\@tempc}}}

\bibitem[\protect\citeauthoryear{{Arcavi} et~al.,}{{Arcavi}
  et~al.}{2014}]{arcavi14}
{Arcavi} I.,  et~al., 2014, \mn@doi [\apj] {10.1088/0004-637X/793/1/38}, \href
  {http://adsabs.harvard.edu/abs/2014ApJ...793...38A} {793, 38}

\bibitem[\protect\citeauthoryear{{Begelman}}{{Begelman}}{1979}]{begelman79}
{Begelman} M.~C.,  1979, \mn@doi [\mnras] {10.1093/mnras/187.2.237}, \href
  {http://adsabs.harvard.edu/abs/1979MNRAS.187..237B} {187, 237}

\bibitem[\protect\citeauthoryear{{Begelman}, {Blandford}  \& {Rees}}{{Begelman}
  et~al.}{1980}]{begelman80}
{Begelman} M.~C.,  {Blandford} R.~D.,   {Rees} M.~J.,  1980, \mn@doi [\nat]
  {10.1038/287307a0}, \href {http://adsabs.harvard.edu/abs/1980Natur.287..307B}
  {287, 307}

\bibitem[\protect\citeauthoryear{{Bellm}}{{Bellm}}{2014}]{bellm14}
{Bellm} E.,  2014, in {Wozniak} P.~R.,  {Graham} M.~J.,  {Mahabal} A.~A.,
  {Seaman} R.,  eds, The Third Hot-wiring the Transient Universe Workshop. pp
  27--33 (\mn@eprint {arXiv} {1410.8185})

\bibitem[\protect\citeauthoryear{{Blagorodnova} et~al.,}{{Blagorodnova}
  et~al.}{2017}]{blagorodnova17}
{Blagorodnova} N.,  et~al., 2017, preprint, \href
  {http://adsabs.harvard.edu/abs/2017arXiv170300965B} {} (\mn@eprint {arXiv}
  {1703.00965})

\bibitem[\protect\citeauthoryear{{Bloom} et~al.,}{{Bloom}
  et~al.}{2011}]{bloom11}
{Bloom} J.~S.,  et~al., 2011, \mn@doi [Science] {10.1126/science.1207150},
  \href {http://adsabs.harvard.edu/abs/2011Sci...333..203B} {333, 203}

\bibitem[\protect\citeauthoryear{{Bonnerot}, {Rossi}, {Lodato}  \&
  {Price}}{{Bonnerot} et~al.}{2016}]{bonnerot16}
{Bonnerot} C.,  {Rossi} E.~M.,  {Lodato} G.,   {Price} D.~J.,  2016, \mn@doi
  [\mnras] {10.1093/mnras/stv2411}, \href
  {http://adsabs.harvard.edu/abs/2016MNRAS.455.2253B} {455, 2253}

\bibitem[\protect\citeauthoryear{{Brown} et~al.,}{{Brown}
  et~al.}{2016}]{brown16}
{Brown} P.~J.,  et~al., 2016, \mn@doi [\apj] {10.3847/0004-637X/828/1/3}, \href
  {http://adsabs.harvard.edu/abs/2016ApJ...828....3B} {828, 3}

\bibitem[\protect\citeauthoryear{{Burrows} et~al.,}{{Burrows}
  et~al.}{2011}]{burrows11}
{Burrows} D.~N.,  et~al., 2011, \mn@doi [\nat] {10.1038/nature10374}, \href
  {http://adsabs.harvard.edu/abs/2011Natur.476..421B} {476, 421}

\bibitem[\protect\citeauthoryear{{Cannizzo}, {Troja}  \& {Lodato}}{{Cannizzo}
  et~al.}{2011}]{cannizzo11}
{Cannizzo} J.~K.,  {Troja} E.,   {Lodato} G.,  2011, \mn@doi [\apj]
  {10.1088/0004-637X/742/1/32}, \href
  {http://adsabs.harvard.edu/abs/2011ApJ...742...32C} {742, 32}

\bibitem[\protect\citeauthoryear{{Chambers} et~al.,}{{Chambers}
  et~al.}{2016}]{chambers16}
{Chambers} K.~C.,  et~al., 2016, preprint, \href
  {http://adsabs.harvard.edu/abs/2016arXiv161205560C} {} (\mn@eprint {arXiv}
  {1612.05560})

\bibitem[\protect\citeauthoryear{{Charisi}, {Bartos}, {Haiman}, {Price-Whelan},
  {Graham}, {Bellm}, {Laher}  \& {M{\'a}rka}}{{Charisi}
  et~al.}{2016}]{charisi16}
{Charisi} M.,  {Bartos} I.,  {Haiman} Z.,  {Price-Whelan} A.~M.,  {Graham}
  M.~J.,  {Bellm} E.~C.,  {Laher} R.~R.,   {M{\'a}rka} S.,  2016, \mn@doi
  [\mnras] {10.1093/mnras/stw1838}, \href
  {http://adsabs.harvard.edu/abs/2016MNRAS.463.2145C} {463, 2145}

\bibitem[\protect\citeauthoryear{{Chatzopoulos}, {Wheeler}, {Vinko}, {Nagy},
  {Wiggins}  \& {Even}}{{Chatzopoulos} et~al.}{2016}]{chatzopoulos16}
{Chatzopoulos} E.,  {Wheeler} J.~C.,  {Vinko} J.,  {Nagy} A.~P.,  {Wiggins}
  B.~K.,   {Even} W.~P.,  2016, \mn@doi [\apj] {10.3847/0004-637X/828/2/94},
  \href {http://adsabs.harvard.edu/abs/2016ApJ...828...94C} {828, 94}

\bibitem[\protect\citeauthoryear{{Chen}, {Madau}, {Sesana}  \& {Liu}}{{Chen}
  et~al.}{2009}]{chen09}
{Chen} X.,  {Madau} P.,  {Sesana} A.,   {Liu} F.~K.,  2009, \mn@doi [\apjl]
  {10.1088/0004-637X/697/2/L149}, \href
  {http://adsabs.harvard.edu/abs/2009ApJ...697L.149C} {697, L149}

\bibitem[\protect\citeauthoryear{{Comerford} et~al.,}{{Comerford}
  et~al.}{2009}]{comerford09}
{Comerford} J.~M.,  et~al., 2009, \mn@doi [\apj] {10.1088/0004-637X/698/1/956},
  \href {http://adsabs.harvard.edu/abs/2009ApJ...698..956C} {698, 956}

\bibitem[\protect\citeauthoryear{{Coughlin} \& {Armitage}}{{Coughlin} \&
  {Armitage}}{2017}]{coughlin17b}
{Coughlin} E.~R.,  {Armitage} P.~J.,  2017, \mn@doi [\mnras]
  {10.1093/mnrasl/slx114}, \href
  {http://adsabs.harvard.edu/abs/2017MNRAS.471L.115C} {471, L115}

\bibitem[\protect\citeauthoryear{{Coughlin} \& {Begelman}}{{Coughlin} \&
  {Begelman}}{2014}]{coughlin14}
{Coughlin} E.~R.,  {Begelman} M.~C.,  2014, \mn@doi [\apj]
  {10.1088/0004-637X/781/2/82}, \href
  {http://adsabs.harvard.edu/abs/2014ApJ...781...82C} {781, 82}

\bibitem[\protect\citeauthoryear{{Coughlin} \& {Begelman}}{{Coughlin} \&
  {Begelman}}{2015}]{coughlin15b}
{Coughlin} E.~R.,  {Begelman} M.~C.,  2015, \mn@doi [\apj]
  {10.1088/0004-637X/809/1/2}, \href
  {http://adsabs.harvard.edu/abs/2015ApJ...809....2C} {809, 2}

\bibitem[\protect\citeauthoryear{{Coughlin} \& {Nixon}}{{Coughlin} \&
  {Nixon}}{2015}]{coughlin15}
{Coughlin} E.~R.,  {Nixon} C.,  2015, \mn@doi [\apjl]
  {10.1088/2041-8205/808/1/L11}, \href
  {http://adsabs.harvard.edu/abs/2015ApJ...808L..11C} {808, L11}

\bibitem[\protect\citeauthoryear{{Coughlin}, {Nixon}, {Begelman}, {Armitage}
  \& {Price}}{{Coughlin} et~al.}{2016}]{coughlin16}
{Coughlin} E.~R.,  {Nixon} C.,  {Begelman} M.~C.,  {Armitage} P.~J.,   {Price}
  D.~J.,  2016, \mn@doi [\mnras] {10.1093/mnras/stv2511}, \href
  {http://adsabs.harvard.edu/abs/2016MNRAS.455.3612C} {455, 3612}

\bibitem[\protect\citeauthoryear{{Coughlin}, {Armitage}, {Nixon}  \&
  {Begelman}}{{Coughlin} et~al.}{2017}]{coughlin17}
{Coughlin} E.~R.,  {Armitage} P.~J.,  {Nixon} C.,   {Begelman} M.~C.,  2017,
  \mn@doi [\mnras] {10.1093/mnras/stw2913}, \href
  {http://adsabs.harvard.edu/abs/2017MNRAS.465.3840C} {465, 3840}

\bibitem[\protect\citeauthoryear{{Cuadra}, {Armitage}, {Alexander}  \&
  {Begelman}}{{Cuadra} et~al.}{2009}]{cuadra09}
{Cuadra} J.,  {Armitage} P.~J.,  {Alexander} R.~D.,   {Begelman} M.~C.,  2009,
  \mn@doi [\mnras] {10.1111/j.1365-2966.2008.14147.x}, \href
  {http://adsabs.harvard.edu/abs/2009MNRAS.393.1423C} {393, 1423}

\bibitem[\protect\citeauthoryear{{D'Orazio} \& {Haiman}}{{D'Orazio} \&
  {Haiman}}{2017}]{dorazio17}
{D'Orazio} D.~J.,  {Haiman} Z.,  2017, preprint, \href
  {http://adsabs.harvard.edu/abs/2017arXiv170201219D} {} (\mn@eprint {arXiv}
  {1702.01219})

\bibitem[\protect\citeauthoryear{{Dong} et~al.,}{{Dong} et~al.}{2016}]{dong16}
{Dong} S.,  et~al., 2016, \mn@doi [Science] {10.1126/science.aac9613}, \href
  {http://adsabs.harvard.edu/abs/2016Sci...351..257D} {351, 257}

\bibitem[\protect\citeauthoryear{{Evans} \& {Kochanek}}{{Evans} \&
  {Kochanek}}{1989}]{evans89}
{Evans} C.~R.,  {Kochanek} C.~S.,  1989, \mn@doi [\apjl] {10.1086/185567},
  \href {http://adsabs.harvard.edu/abs/1989ApJ...346L..13E} {346, L13}

\bibitem[\protect\citeauthoryear{{Frank} \& {Rees}}{{Frank} \&
  {Rees}}{1976}]{frank76}
{Frank} J.,  {Rees} M.~J.,  1976, \mn@doi [\mnras] {10.1093/mnras/176.3.633},
  \href {http://adsabs.harvard.edu/abs/1976MNRAS.176..633F} {176, 633}

\bibitem[\protect\citeauthoryear{{Gafton} \& {Rosswog}}{{Gafton} \&
  {Rosswog}}{2011}]{gafton11}
{Gafton} E.,  {Rosswog} S.,  2011, \mn@doi [\mnras]
  {10.1111/j.1365-2966.2011.19528.x}, \href
  {http://adsabs.harvard.edu/abs/2011MNRAS.418..770G} {418, 770}

\bibitem[\protect\citeauthoryear{{Gezari} et~al.,}{{Gezari}
  et~al.}{2012}]{gezari12}
{Gezari} S.,  et~al., 2012, \mn@doi [\nat] {10.1038/nature10990}, \href
  {http://adsabs.harvard.edu/abs/2012Natur.485..217G} {485, 217}

\bibitem[\protect\citeauthoryear{{Giannios} \& {Metzger}}{{Giannios} \&
  {Metzger}}{2011}]{giannios11}
{Giannios} D.,  {Metzger} B.~D.,  2011, \mn@doi [\mnras]
  {10.1111/j.1365-2966.2011.19188.x}, \href
  {http://adsabs.harvard.edu/abs/2011MNRAS.416.2102G} {416, 2102}

\bibitem[\protect\citeauthoryear{{Godoy-Rivera} et~al.,}{{Godoy-Rivera}
  et~al.}{2017}]{godoy17}
{Godoy-Rivera} D.,  et~al., 2017, \mn@doi [\mnras] {10.1093/mnras/stw3237},
  \href {http://adsabs.harvard.edu/abs/2017MNRAS.466.1428G} {466, 1428}

\bibitem[\protect\citeauthoryear{{Guillochon} \& {Ramirez-Ruiz}}{{Guillochon}
  \& {Ramirez-Ruiz}}{2013}]{guillochon13}
{Guillochon} J.,  {Ramirez-Ruiz} E.,  2013, \mn@doi [\apj]
  {10.1088/0004-637X/767/1/25}, \href
  {http://adsabs.harvard.edu/abs/2013ApJ...767...25G} {767, 25}

\bibitem[\protect\citeauthoryear{{Guillochon} \& {Ramirez-Ruiz}}{{Guillochon}
  \& {Ramirez-Ruiz}}{2015}]{guillochon15}
{Guillochon} J.,  {Ramirez-Ruiz} E.,  2015, \mn@doi [\apj]
  {10.1088/0004-637X/809/2/166}, \href
  {http://adsabs.harvard.edu/abs/2015ApJ...809..166G} {809, 166}

\bibitem[\protect\citeauthoryear{{Guillochon}, {Manukian}  \&
  {Ramirez-Ruiz}}{{Guillochon} et~al.}{2014}]{guillochon14}
{Guillochon} J.,  {Manukian} H.,   {Ramirez-Ruiz} E.,  2014, \mn@doi [\apj]
  {10.1088/0004-637X/783/1/23}, \href
  {http://adsabs.harvard.edu/abs/2014ApJ...783...23G} {783, 23}

\bibitem[\protect\citeauthoryear{{Guillochon}, {McCourt}, {Chen}, {Johnson}  \&
  {Berger}}{{Guillochon} et~al.}{2016}]{guillochon16}
{Guillochon} J.,  {McCourt} M.,  {Chen} X.,  {Johnson} M.~D.,   {Berger} E.,
  2016, \mn@doi [\apj] {10.3847/0004-637X/822/1/48}, \href
  {http://adsabs.harvard.edu/abs/2016ApJ...822...48G} {822, 48}

\bibitem[\protect\citeauthoryear{{Hayasaki}, {Mineshige}  \&
  {Sudou}}{{Hayasaki} et~al.}{2007}]{hayasaki07}
{Hayasaki} K.,  {Mineshige} S.,   {Sudou} H.,  2007, \mn@doi [\pasj]
  {10.1093/pasj/59.2.427}, \href
  {http://adsabs.harvard.edu/abs/2007PASJ...59..427H} {59, 427}

\bibitem[\protect\citeauthoryear{{Hayasaki}, {Stone}  \& {Loeb}}{{Hayasaki}
  et~al.}{2013}]{hayasaki13}
{Hayasaki} K.,  {Stone} N.,   {Loeb} A.,  2013, \mn@doi [\mnras]
  {10.1093/mnras/stt871}, \href
  {http://adsabs.harvard.edu/abs/2013MNRAS.434..909H} {434, 909}

\bibitem[\protect\citeauthoryear{{Hayasaki}, {Stone}  \& {Loeb}}{{Hayasaki}
  et~al.}{2016}]{hayasaki16}
{Hayasaki} K.,  {Stone} N.,   {Loeb} A.,  2016, \mn@doi [\mnras]
  {10.1093/mnras/stw1387}, \href
  {http://adsabs.harvard.edu/abs/2016MNRAS.461.3760H} {461, 3760}

\bibitem[\protect\citeauthoryear{{Holoien} et~al.,}{{Holoien}
  et~al.}{2014}]{holoien14}
{Holoien} T.~W.-S.,  et~al., 2014, \mn@doi [\mnras] {10.1093/mnras/stu1922},
  \href {http://adsabs.harvard.edu/abs/2014MNRAS.445.3263H} {445, 3263}

\bibitem[\protect\citeauthoryear{{Ivanov}, {Polnarev}  \& {Saha}}{{Ivanov}
  et~al.}{2005}]{ivanov05}
{Ivanov} P.~B.,  {Polnarev} A.~G.,   {Saha} P.,  2005, \mn@doi [\mnras]
  {10.1111/j.1365-2966.2005.08843.x}, \href
  {http://adsabs.harvard.edu/abs/2005MNRAS.358.1361I} {358, 1361}

\bibitem[\protect\citeauthoryear{{Ivezic} et~al.,}{{Ivezic}
  et~al.}{2008}]{ivezic08}
{Ivezic} Z.,  et~al., 2008, preprint, \href
  {http://adsabs.harvard.edu/abs/2008arXiv0805.2366I} {} (\mn@eprint {arXiv}
  {0805.2366})

\bibitem[\protect\citeauthoryear{{Kaiser} et~al.,}{{Kaiser}
  et~al.}{2010}]{kaiser10}
{Kaiser} N.,  et~al., 2010, in Ground-based and Airborne Telescopes III. p.
  77330E, \mn@doi{10.1117/12.859188}

\bibitem[\protect\citeauthoryear{{Kelley}, {Blecha}  \& {Hernquist}}{{Kelley}
  et~al.}{2017}]{kelley17}
{Kelley} L.~Z.,  {Blecha} L.,   {Hernquist} L.,  2017, \mn@doi [\mnras]
  {10.1093/mnras/stw2452}, \href
  {http://adsabs.harvard.edu/abs/2017MNRAS.464.3131K} {464, 3131}

\bibitem[\protect\citeauthoryear{{Kesden}}{{Kesden}}{2012}]{kesden12}
{Kesden} M.,  2012, \mn@doi [\prd] {10.1103/PhysRevD.85.024037}, \href
  {http://adsabs.harvard.edu/abs/2012PhRvD..85b4037K} {85, 024037}

\bibitem[\protect\citeauthoryear{{Komossa}}{{Komossa}}{2015}]{komossa15}
{Komossa} S.,  2015, \mn@doi [Journal of High Energy Astrophysics]
  {10.1016/j.jheap.2015.04.006}, \href
  {http://adsabs.harvard.edu/abs/2015JHEAp...7..148K} {7, 148}

\bibitem[\protect\citeauthoryear{{Komossa}, {Burwitz}, {Hasinger}, {Predehl},
  {Kaastra}  \& {Ikebe}}{{Komossa} et~al.}{2003}]{komossa03}
{Komossa} S.,  {Burwitz} V.,  {Hasinger} G.,  {Predehl} P.,  {Kaastra} J.~S.,
  {Ikebe} Y.,  2003, \mn@doi [\apjl] {10.1086/346145}, \href
  {http://adsabs.harvard.edu/abs/2003ApJ...582L..15K} {582, L15}

\bibitem[\protect\citeauthoryear{{Lacy}, {Townes}  \& {Hollenbach}}{{Lacy}
  et~al.}{1982}]{lacy82}
{Lacy} J.~H.,  {Townes} C.~H.,   {Hollenbach} D.~J.,  1982, \mn@doi [\apj]
  {10.1086/160402}, \href {http://adsabs.harvard.edu/abs/1982ApJ...262..120L}
  {262, 120}

\bibitem[\protect\citeauthoryear{{Lauer} et~al.,}{{Lauer}
  et~al.}{2005}]{lauer05}
{Lauer} T.~R.,  et~al., 2005, \mn@doi [\aj] {10.1086/429565}, \href
  {http://adsabs.harvard.edu/abs/2005AJ....129.2138L} {129, 2138}

\bibitem[\protect\citeauthoryear{{Law} et~al.,}{{Law} et~al.}{2009}]{law09}
{Law} N.~M.,  et~al., 2009, \mn@doi [\pasp] {10.1086/648598}, \href
  {http://adsabs.harvard.edu/abs/2009PASP..121.1395L} {121, 1395}

\bibitem[\protect\citeauthoryear{{Leloudas} et~al.,}{{Leloudas}
  et~al.}{2016}]{leloudas16}
{Leloudas} G.,  et~al., 2016, \mn@doi [Nature Astronomy]
  {10.1038/s41550-016-0002}, \href
  {http://adsabs.harvard.edu/abs/2016NatAs...1E...2L} {1, 0002}

\bibitem[\protect\citeauthoryear{{Levan} et~al.,}{{Levan}
  et~al.}{2011}]{levan11}
{Levan} A.~J.,  et~al., 2011, \mn@doi [Science] {10.1126/science.1207143},
  \href {http://adsabs.harvard.edu/abs/2011Sci...333..199L} {333, 199}

\bibitem[\protect\citeauthoryear{{Li}, {Naoz}, {Kocsis}  \& {Loeb}}{{Li}
  et~al.}{2015}]{li15}
{Li} G.,  {Naoz} S.,  {Kocsis} B.,   {Loeb} A.,  2015, \mn@doi [\mnras]
  {10.1093/mnras/stv1031}, \href
  {http://adsabs.harvard.edu/abs/2015MNRAS.451.1341L} {451, 1341}

\bibitem[\protect\citeauthoryear{{Liu}, {Li}  \& {Chen}}{{Liu}
  et~al.}{2009}]{liu09}
{Liu} F.~K.,  {Li} S.,   {Chen} X.,  2009, \mn@doi [\apjl]
  {10.1088/0004-637X/706/1/L133}, \href
  {http://adsabs.harvard.edu/abs/2009ApJ...706L.133L} {706, L133}

\bibitem[\protect\citeauthoryear{{Liu} et~al.,}{{Liu} et~al.}{2016}]{liu16}
{Liu} T.,  et~al., 2016, \mn@doi [\apj] {10.3847/0004-637X/833/1/6}, \href
  {http://adsabs.harvard.edu/abs/2016ApJ...833....6L} {833, 6}

\bibitem[\protect\citeauthoryear{{Lodato}, {King}  \& {Pringle}}{{Lodato}
  et~al.}{2009}]{lodato09}
{Lodato} G.,  {King} A.~R.,   {Pringle} J.~E.,  2009, \mn@doi [\mnras]
  {10.1111/j.1365-2966.2008.14049.x}, \href
  {http://adsabs.harvard.edu/abs/2009MNRAS.392..332L} {392, 332}

\bibitem[\protect\citeauthoryear{{Loeb} \& {Ulmer}}{{Loeb} \&
  {Ulmer}}{1997}]{loeb97}
{Loeb} A.,  {Ulmer} A.,  1997, \mn@doi [\apj] {10.1086/304814}, \href
  {http://adsabs.harvard.edu/abs/1997ApJ...489..573L} {489, 573}

\bibitem[\protect\citeauthoryear{{MacFadyen} \&
  {Milosavljevi{\'c}}}{{MacFadyen} \& {Milosavljevi{\'c}}}{2008}]{macfadyen08}
{MacFadyen} A.~I.,  {Milosavljevi{\'c}} M.,  2008, \mn@doi [\apj]
  {10.1086/523869}, \href {http://adsabs.harvard.edu/abs/2008ApJ...672...83M}
  {672, 83}

\bibitem[\protect\citeauthoryear{{Madigan}, {Halle}, {Moody}, {McCourt}  \&
  {Nixon}}{{Madigan} et~al.}{2017}]{madigan17}
{Madigan} A.-M.,  {Halle} A.,  {Moody} M.,  {McCourt} M.,   {Nixon} C.,  2017,
  preprint, \href {http://adsabs.harvard.edu/abs/2017arXiv170503462M} {}
  (\mn@eprint {arXiv} {1705.03462})

\bibitem[\protect\citeauthoryear{{Mainetti}, {Lupi}, {Campana}, {Colpi},
  {Coughlin}, {Guillochon}  \& {Ramirez-Ruiz}}{{Mainetti}
  et~al.}{2017}]{mainetti17}
{Mainetti} D.,  {Lupi} A.,  {Campana} S.,  {Colpi} M.,  {Coughlin} E.~R.,
  {Guillochon} J.,   {Ramirez-Ruiz} E.,  2017, \mn@doi [\aap]
  {10.1051/0004-6361/201630092}, \href
  {http://adsabs.harvard.edu/abs/2017A%26A...600A.124M} {600, A124}

\bibitem[\protect\citeauthoryear{{Margutti} et~al.,}{{Margutti}
  et~al.}{2017}]{margutti17}
{Margutti} R.,  et~al., 2017, \mn@doi [\apj] {10.3847/1538-4357/836/1/25},
  \href {http://adsabs.harvard.edu/abs/2017ApJ...836...25M} {836, 25}

\bibitem[\protect\citeauthoryear{{McConnell} \& {Ma}}{{McConnell} \&
  {Ma}}{2013}]{mcconnell13}
{McConnell} N.~J.,  {Ma} C.-P.,  2013, \mn@doi [\apj]
  {10.1088/0004-637X/764/2/184}, \href
  {http://adsabs.harvard.edu/abs/2013ApJ...764..184M} {764, 184}

\bibitem[\protect\citeauthoryear{{Metzger} \& {Stone}}{{Metzger} \&
  {Stone}}{2016}]{metzger16}
{Metzger} B.~D.,  {Stone} N.~C.,  2016, \mn@doi [\mnras]
  {10.1093/mnras/stw1394}, \href
  {http://adsabs.harvard.edu/abs/2016MNRAS.461..948M} {461, 948}

\bibitem[\protect\citeauthoryear{{Neill} et~al.,}{{Neill}
  et~al.}{2011}]{neill11}
{Neill} J.~D.,  et~al., 2011, \mn@doi [\apj] {10.1088/0004-637X/727/1/15},
  \href {http://adsabs.harvard.edu/abs/2011ApJ...727...15N} {727, 15}

\bibitem[\protect\citeauthoryear{Peters \& Mathews}{Peters \&
  Mathews}{1963}]{peters63}
Peters P.~C.,  Mathews J.,  1963, \mn@doi [Phys. Rev.]
  {10.1103/PhysRev.131.435}, 131, 435

\bibitem[\protect\citeauthoryear{{Phinney}}{{Phinney}}{1989}]{phinney89}
{Phinney} E.~S.,  1989, in {Morris} M.,  ed.,  IAU Symposium Vol. 136, The
  Center of the Galaxy. p.~543

\bibitem[\protect\citeauthoryear{{Price}}{{Price}}{2007}]{price07}
{Price} D.~J.,  2007, \mn@doi [\pasa] {10.1071/AS07022}, \href
  {http://adsabs.harvard.edu/abs/2007PASA...24..159P} {24, 159}

\bibitem[\protect\citeauthoryear{{Price} et~al.,}{{Price}
  et~al.}{2017}]{price17}
{Price} D.~J.,  et~al., 2017, preprint, \href
  {http://adsabs.harvard.edu/abs/2017arXiv170203930P} {} (\mn@eprint {arXiv}
  {1702.03930})

\bibitem[\protect\citeauthoryear{{Quimby} et~al.,}{{Quimby}
  et~al.}{2011}]{quimby11}
{Quimby} R.~M.,  et~al., 2011, \mn@doi [\nat] {10.1038/nature10095}, \href
  {http://adsabs.harvard.edu/abs/2011Natur.474..487Q} {474, 487}

\bibitem[\protect\citeauthoryear{{Rees}}{{Rees}}{1988}]{rees88}
{Rees} M.~J.,  1988, \mn@doi [\nat] {10.1038/333523a0}, \href
  {http://adsabs.harvard.edu/abs/1988Natur.333..523R} {333, 523}

\bibitem[\protect\citeauthoryear{{Ricarte}, {Natarajan}, {Dai}  \&
  {Coppi}}{{Ricarte} et~al.}{2016}]{ricarte16}
{Ricarte} A.,  {Natarajan} P.,  {Dai} L.,   {Coppi} P.,  2016, \mn@doi [\mnras]
  {10.1093/mnras/stw355}, \href
  {http://adsabs.harvard.edu/abs/2016MNRAS.458.1712R} {458, 1712}

\bibitem[\protect\citeauthoryear{{Rodriguez}, {Taylor}, {Zavala}, {Peck},
  {Pollack}  \& {Romani}}{{Rodriguez} et~al.}{2006}]{rodriguez06}
{Rodriguez} C.,  {Taylor} G.~B.,  {Zavala} R.~T.,  {Peck} A.~B.,  {Pollack}
  L.~K.,   {Romani} R.~W.,  2006, \mn@doi [\apj] {10.1086/504825}, \href
  {http://adsabs.harvard.edu/abs/2006ApJ...646...49R} {646, 49}

\bibitem[\protect\citeauthoryear{{Roth}, {Kasen}, {Guillochon}  \&
  {Ramirez-Ruiz}}{{Roth} et~al.}{2016}]{roth16}
{Roth} N.,  {Kasen} D.,  {Guillochon} J.,   {Ramirez-Ruiz} E.,  2016, \mn@doi
  [\apj] {10.3847/0004-637X/827/1/3}, \href
  {http://adsabs.harvard.edu/abs/2016ApJ...827....3R} {827, 3}

\bibitem[\protect\citeauthoryear{{S{\c a}dowski}, {Tejeda}, {Gafton}, {Rosswog}
   \& {Abarca}}{{S{\c a}dowski} et~al.}{2016}]{sadowski16}
{S{\c a}dowski} A.,  {Tejeda} E.,  {Gafton} E.,  {Rosswog} S.,   {Abarca} D.,
  2016, \mn@doi [\mnras] {10.1093/mnras/stw589}, \href
  {http://adsabs.harvard.edu/abs/2016MNRAS.458.4250S} {458, 4250}

\bibitem[\protect\citeauthoryear{{Shakura} \& {Sunyaev}}{{Shakura} \&
  {Sunyaev}}{1973}]{shakura73}
{Shakura} N.~I.,  {Sunyaev} R.~A.,  1973, \aap, \href
  {http://adsabs.harvard.edu/abs/1973A%26A....24..337S} {24, 337}

\bibitem[\protect\citeauthoryear{{Shappee} et~al.,}{{Shappee}
  et~al.}{2014}]{shappee14}
{Shappee} B.~J.,  et~al., 2014, \mn@doi [\apj] {10.1088/0004-637X/788/1/48},
  \href {http://adsabs.harvard.edu/abs/2014ApJ...788...48S} {788, 48}

\bibitem[\protect\citeauthoryear{{Shiokawa}, {Krolik}, {Cheng}, {Piran}  \&
  {Noble}}{{Shiokawa} et~al.}{2015}]{shiokawa15}
{Shiokawa} H.,  {Krolik} J.~H.,  {Cheng} R.~M.,  {Piran} T.,   {Noble} S.~C.,
  2015, \mn@doi [\apj] {10.1088/0004-637X/804/2/85}, \href
  {http://adsabs.harvard.edu/abs/2015ApJ...804...85S} {804, 85}

\bibitem[\protect\citeauthoryear{{Stone} \& {Metzger}}{{Stone} \&
  {Metzger}}{2016}]{stone16}
{Stone} N.~C.,  {Metzger} B.~D.,  2016, \mn@doi [\mnras]
  {10.1093/mnras/stv2281}, \href
  {http://adsabs.harvard.edu/abs/2016MNRAS.455..859S} {455, 859}

\bibitem[\protect\citeauthoryear{{Stone}, {Sari}  \& {Loeb}}{{Stone}
  et~al.}{2013}]{stone13}
{Stone} N.,  {Sari} R.,   {Loeb} A.,  2013, \mn@doi [\mnras]
  {10.1093/mnras/stt1270}, \href
  {http://adsabs.harvard.edu/abs/2013MNRAS.435.1809S} {435, 1809}

\bibitem[\protect\citeauthoryear{{Strubbe} \& {Quataert}}{{Strubbe} \&
  {Quataert}}{2009}]{strubbe09}
{Strubbe} L.~E.,  {Quataert} E.,  2009, \mn@doi [\mnras]
  {10.1111/j.1365-2966.2009.15599.x}, \href
  {http://adsabs.harvard.edu/abs/2009MNRAS.400.2070S} {400, 2070}

\bibitem[\protect\citeauthoryear{{Tchekhovskoy}, {Metzger}, {Giannios}  \&
  {Kelley}}{{Tchekhovskoy} et~al.}{2014}]{tchekhovskoy14}
{Tchekhovskoy} A.,  {Metzger} B.~D.,  {Giannios} D.,   {Kelley} L.~Z.,  2014,
  \mn@doi [\mnras] {10.1093/mnras/stt2085}, \href
  {http://adsabs.harvard.edu/abs/2014MNRAS.437.2744T} {437, 2744}

\bibitem[\protect\citeauthoryear{{Zauderer} et~al.,}{{Zauderer}
  et~al.}{2011}]{zauderer11}
{Zauderer} B.~A.,  et~al., 2011, \mn@doi [\nat] {10.1038/nature10366}, \href
  {http://adsabs.harvard.edu/abs/2011Natur.476..425Z} {476, 425}

\makeatother
\end{thebibliography}

\bsp	
\label{lastpage}
\end{document}